\documentclass[review]{elsarticle}

\usepackage{lineno,hyperref}

\usepackage{graphicx,color}
\usepackage{subfig}
\usepackage{dcolumn}
\usepackage{multirow}
\usepackage{bm}
\usepackage{amsmath}
\usepackage{mathtools}
\usepackage{cancel}
\usepackage{adjustbox}

\modulolinenumbers[5]

\journal{Journal of \LaTeX\ Templates}

\begin{document}

\begin{frontmatter}
\title{The role of the screening potential in the deuteron-deuteron thermonuclear reaction rates}

\author[birjand]{Faisal Etminan\corref{mycorrespondingauthor}}
\cortext[mycorrespondingauthor]{Corresponding author}
\ead{fetminan@birjand.ac.ir}
%

%

%
\address[birjand]{
		Department of Physics, Faculty of Sciences, University of Birjand, Birjand 97175-615, Iran
}
\begin{abstract}
		The deuteron-deuteron (D-D) thermonuclear reaction rates in metallic environments, accounting for electron screening effects, are calculated using S-factor functions derived from fits to low-energy D-D reaction data. For this purpose, a fitted S-factor model based on the NACRE compilation is employed, which constrains the energy range applicable to Big Bang nucleosynthesis (BBN) for the $^{2}\textrm{H}\left(d,p\right)^{3}\textrm{H}$ and $^{2}\textrm{H}\left(d,n\right)^{3}\textrm{He}$ reactions.  
The Maxwellian-averaged thermonuclear reaction rates, relevant to astrophysical plasmas at temperatures ranging from $10^{6}$ K to $10^{10}$ K (or $1.3 \times 10^{8}$ K), are presented in tabular formats. 
The effects of electron screening are phenomenologically analyzed, with screening energy ($U_{e}$) values of 100, 400, 750, 1000, and 1250 eV being employed for this purpose.
 This selection of values is grounded in theoretical and experimental studies conducted to date. 
Ultimately, the numerical analysis reveals that the ratio of the screened reaction rate to the unscreened reaction rate can be expressed by the numerical formula  $ \exp\left(4.70 +6.50{\times10^{-6}}U_{e}/{T_{9}}\right) $
for both the $^{2}\textrm{H}\left(d,p\right)^{3}\textrm{H}$ and $^{2}\textrm{H}\left(d,n\right)^{3}\textrm{He}$ reactions.  
\end{abstract}


\end{frontmatter}

	\section{Introduction} \label{sec:intro}
	In astrophysical environments and contemporary fusion experiments, nuclear reactions are observed to occur in the presence of other charges, which influence the rates at which these reactions proceed. Consequently, the fusion cross-sections and Maxwellian reactivities must be determined as precisely as possible to interpret fusion rate measurements and predict the performance of future fusion devices~\cite{Bosch-NF-1992, IchimaruRevModPhys}. 

In a plasma environment, a balance is struck between the Coulomb potential and thermal (kinetic) energy~\cite{PhysRevA.29.945,AliottaFPhy2022}. 
At astrophysical energies, well below the Coulomb barrier, the electron screening effect can account for the exponential-like enhancement of nuclear fusion cross-sections measured in atomic environments compared to bare nuclei media~\cite{PhysRevC.83.018801}. If the most effective energy for nuclear reactions ($E_{0}$ in Eq.~\eqref{eq:gamow-energy}, also known as the Gamow peak energy) significantly exceeds the thermal energy, density fluctuations can activate fusion reactions even at $ 0 \, \text{K}$~\cite{ANGULO19993}. This enhancement was first observed in the $^{2}\textrm{H}\left(d,p\right)^{3}\textrm{H}$ and $^{2}\textrm{H}\left(d,n\right)^{3}\textrm{He}$ reactions~\cite{Czerski_EPL2001} and later confirmed by numerous experiments~\cite{RAIOLA2002193,raiola2002enhanced,Jirohta2004,CRUZ2005181} involving deuteron beams and deuterated targets~\cite{BONOMO2003C37,greife1995oppenheimer,Schenkel2019}.

Under terrestrial conditions, the D-D reactions in metallic environments serve as an ideal tool due to the relatively significant $S$ factor of this reaction. The presence of quasi-free conduction and shell electrons in the metal lattice can lead to substantial enhancements in reaction yields~\cite{HukePRC2008, CzerskiPRC2022, Berlinguette2019RevisitingTC, chen2025electrochemical}.  
However, the practical feasibility of achieving thermonuclear fusion conditions on Earth that leverage this electron screening effect remains uncertain. If feasible, the extent of this effect's contribution to reaction rates must be quantified. %

In light of these considerations, in this work, the D-D Maxwellian-averaged thermonuclear reaction rates are reevaluated, incorporating screening effects to simulate metallic deuterium plasma environments.
Although Ref.~\cite{Tumino_2014} offers the most current reaction rate calculations, a fitted $S$-factor model based on the NACRE (Nuclear Astrophysics Compilation of REactions) database~\cite{ANGULO19993, XU201361} was employed in this study. The NACRE compilation, updated in Ref.~\cite{XU201361}, incorporates extensive experimental data and aligns well with measurements from Ref.~\cite{Tumino_2014}. However, its applicability is limited to the BBN energy range for the $^{2}\textrm{H}\left(d,p\right)^{3}\textrm{H}$ and $^{2}\textrm{H}\left(d,n\right)^{3}\textrm{He}$ reactions. Moreover, Coc et al.~\cite{cocPRD2015} recently reassessed the D-D reactions rates governing deuterium destruction.

Additionally, the numerical results for reaction rates, both with and without screening effects, are presented in tabular form as functions of temperature. The dependence of the reaction rate on the screening potential, $U_{e}$, is investigated by calculating the ratio of the screened to unscreened reaction rates as a function of $U_{e} / (k_{B}T)$, where $k_{B}$ is the Boltzmann constant. These results are fitted to an analytical function, providing insights for optimizing experimental designs to achieve fusion. 
It is emphasized that the effects of electron screening are described phenomenologically in this work, as opposed to being addressed through a novel theoretical framework.

The paper is structured as follows:  
In Section~\ref{sec:general-formula}, the formalism adopted for deriving the astrophysical rates of charged particle-induced reactions is described. This section includes the general formulae used for calculating the Maxwellian-averaged reaction rates, along with a discussion of the screened nuclear reaction cross-section and the non-resonant NACRE model.  
The numerical results are presented in Section~\ref{sec:numerical-result}.  
A summary and conclusions are given in Section~\ref{sec:Summary-and-conclusions}.  

\section{General formalism}  \label{sec:general-formula}
The reaction rates formalism has been explained in detail in various publications such as~\cite{ANGULO19993,fowler1967,DESCOUVEMONT2004203}. Here, only the essential formulae are provided.
\subsection{Maxwellian-averaged reaction rates $N_{A}\left\langle \sigma v\right\rangle $}
For two-body reactions characterized by a relative velocity $ v $ and a cross-section $ \sigma $ (in barn), the Maxwellian-averaged reaction rates $ N_{A}\left\langle \sigma v\right\rangle $ are defined in  units of $\textrm{cm}^{3} \textrm{mol}^{-1} \textrm{ s}^{-1}$~\cite{fowler1967},  
\begin{equation}  
	N_{A}\left\langle \sigma v\right\rangle = 3.7313 \times 10^{10} A^{-1/2} T_{9}^{-3/2} \int_{0}^{\infty} \sigma\left(E\right) E \exp\left(-11.605 \: E / T_{9}\right) dE, \label{eq:reaction-rate}  
\end{equation}  
where $ N_{A} $ is the Avogadro number, $ A $ the reduced mass in atomic mass units (amu), $ E $  energy (in MeV) in the center-of-mass (CM) reference frame and $ T_{9} $ represents the temperature in units of $ 10^{9} $ K. The numerical calculation of the rates is performed over the temperature range $ T_{9} = 0.001 $ to $ T_{9} = 10 $.

For nuclear fusion reactions occurring at thermal energies $ E $ significantly below the Coulomb barrier between interacting nuclei, the nuclear fusion cross-section of a bare nucleus $ \sigma_{\textrm{bare}} $ is typically expressed as~\cite{RevModPhys.29.547}  
\begin{equation}  
	\sigma_{\textrm{bare}}\left(E\right) = \frac{S\left(E\right)}{E} \exp\left(-2\pi\eta\right), \label{eq:bare-cross-section}  
\end{equation}  
where $ S\left(E\right) $, known as the astrophysical $ S $-factor, accounts for nuclear effects and is generally a slowly varying function of energy. Additionally, it factors out the energy dependence of the de Broglie wavelength and the Coulomb penetrability from $ \sigma\left(E\right) $.   
The Sommerfeld parameter $ \eta $ is given by $ \eta = Z_{1}Z_{2}e^{2}/\hbar v = 0.1575 \: Z_{1}Z_{2}\left(\sqrt{A/E}\right) $, where $ Z_{1} $ and $ Z_{2} $ are the charge numbers of the interacting nuclei, and $ \hbar $ is the reduced Planck constant.  

The above formulations do not account for screening effects caused by electrons bound to the molecules of the target. However, these effects must eventually be considered as the energy decreases. A modified version of the cross-section for lower energies will be introduced in Subsection~\ref{subsec:non-resonant}.

\begin{figure*}[htp]
	\centering
	\includegraphics[scale=0.8]{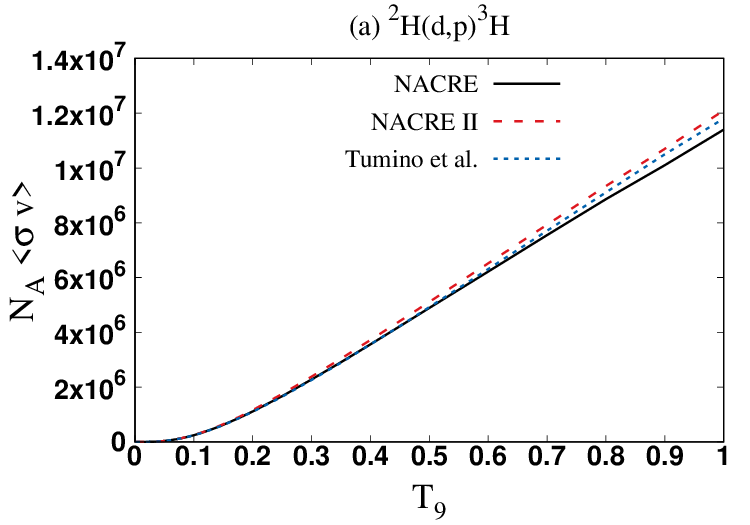} \includegraphics[scale=0.8]{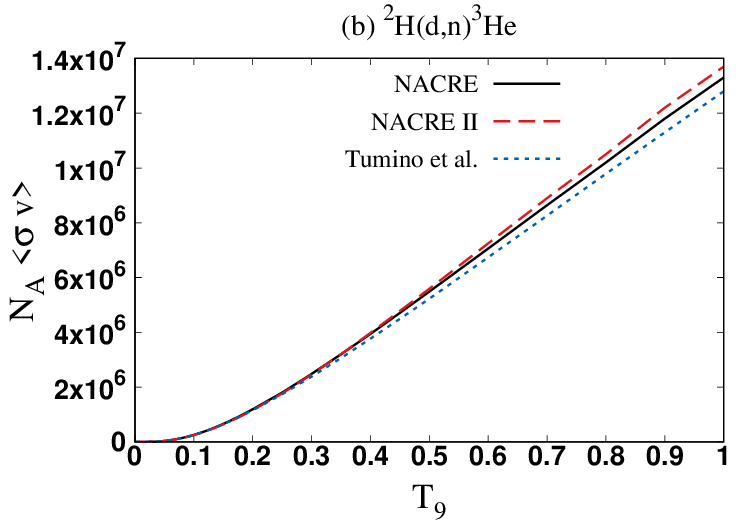} 
	\caption{
		The reaction rates, \(N_{A}\left\langle \sigma v\right\rangle\) in Eq.~\eqref{eq:reaction-rate}, with \(U_{e}=0\) are compared for original NACRE by Angulo et al.~\cite{ANGULO19993} (solid black line), NACRE II by Xu et al. ~\cite{XU201361} (dashed red line) and Tumino et al.~\cite{Tumino_2014} (dotted blue line) for the reactions (a) \(^{2}\textrm{H}(d,p)^{3}\textrm{H}\) and (b) \(^{2}\textrm{H}(d,n)^{3}\textrm{He}\) along \(T_{9}\).
		\label{fig:compare-rate}}	
\end{figure*}

To justify the choice of the original NACRE methodology (Angulo et al.~\cite{ANGULO19993}), the Fig.~\ref{fig:compare-rate} is provided in which the reaction rates, \(N_{A}\left\langle \sigma v\right\rangle\) in Eq.~\eqref{eq:reaction-rate}, with \(U_{e}=0\) are compared for the original NACRE by Angulo et al.~\cite{ANGULO19993}, NACRE II compilation by Xu et al.~\cite{XU201361} and Tumino et al. ~\cite{Tumino_2014}.
This could also provide a basic error estimate for the \(\left\langle \sigma v\right\rangle\) fitting, which will be performed in Section~\ref{sec:numerical-result}, as demonstrated in Fig.~\ref{fig:ddp-ReaRat-fit}.

\subsection{Numerical integration of the rates}
Except around narrow resonances, the $S$-factor is observed to be a smooth function of energy, making it suitable for the extrapolation of measured cross-sections down to astrophysical energies~\cite{ANGULO19993,DESCOUVEMONT2004203,BrownPRC1391}. When $S\left(E\right)$ is assumed to be constant, the integrand in Eq.~\eqref{eq:reaction-rate} is found to peak at the "most effective energy" (also referred to as the Gamow energy)~\cite{ANGULO19993,fowler1967,DESCOUVEMONT2004203,pinesPRC2020}, given by  
\begin{eqnarray}  
	E_{0} & = & \left(\frac{\mu}{2}\right)^{1/3}\left(\frac{\pi e^{2}Z_{1}Z_{2}\:k_{B}T}{\hbar}\right)^{2/3}=0.1220\left(Z_{1}^{2}Z_{2}^{2}A\right)^{1/3}T_{9}^{2/3},\label{eq:gamow-energy}  
\end{eqnarray}  
 where $ \mu $ denotes the reduced mass of the system. This peak can be approximated by a Gaussian function centered at $E_{0}$, with the full width at $1/e$ of the maximum expressed as  
\begin{eqnarray}  
	\Delta E_{0}&=&4\left(\frac{E_{0}k_{B}T}{3}\right)^{1/2}=0.2368\left(Z_{1}^{2}Z_{2}^{2}A\right)^{1/6}T_{9}^{5/6}.\label{eq:DeltaE-FWM}  
\end{eqnarray}  

Under these assumptions, the integral in Eq.~\eqref{eq:reaction-rate} can be calculated analytically~\cite{ANGULO19993,fowler1967,DESCOUVEMONT2004203}. However, in this work, these approximations were not solely relied upon; instead, the integral was evaluated numerically for the non-resonant contribution to the reaction rate. For a given temperature, the integration was performed over the energy domain $\left(E_{0}-n^{\prime}\:\Delta E_{0}, E_{0}+n^{\prime}\:\Delta E_{0}\right)$, with $n^{\prime}=2$ (as in Refs.~\cite{ANGULO19993,DESCOUVEMONT2004203}), using the 16-point Gaussian quadrature rule.  

\subsection{Electron screening by fusion plasma particles}\label{subsec:electronScreen}
Essentially, plasmas studied in fusion research are quasi-neutral, 
but there is a characteristic scale length for shielding the potential
due to individual charges, 
that is known as the Debye screening length, and depends equally on the plasma\textquoteright s temperature and density. 
Since the mass ratio $m_{D}/m_{e}\gg1$, electrons respond to the perturbations much faster than deuterons (ions), 
and the ions can be considered as background stationary particles to some degree, i.e., 
the effective density $n_{e}$ of the free charge carriers (electrons) forms a Debye sphere $\lambda_{De}$ around the deuterons, which leads to generating the screening potential. 
Therefore, the characteristic length of plasma for shielding, Debye screening length is determined by 
\begin{equation}
	\text{\ensuremath{\lambda_{De}=\left(\frac{k_{B}T_{e}}{4\pi n_{e}e^{2}}\right)^{1/2}.}} \label{eq:Debye}
\end{equation}
Also, the electron screening, or, to be more accurate, the plasma-particle screening potential energy $U_{e}$ can be written (as given in Ref.~\cite{pinesPRC2020}), 
\begin{equation}
	U_{e}=\frac{e^{2}}{\lambda_{De}}. \label{eq:Ue}
\end{equation}
Similarly, the electron density (in ${cm^{-3}}$) regarding temperature and screening potential is
\begin{equation}
	n_{e}=U_{e}^{2} \: \frac{0.08617}{4\pi e^{6}}\:T_{9}\times10^{27} \simeq 2.3\times10^{24} \: U_{e}^{2}\:T_{9}. \label{eq:neVsTe}
\end{equation}

In Ref.~\cite{Schenkel2019} Schenkel et al. reported results for D-D fusion with palladium wires where D-D fusion reactions are studied in a pulsed plasma in the glow discharge regime using a benchtop apparatus. 
They found neutron yields as a function of cathode voltage over $100$ times higher than yields expected for bare nuclei fusion at ion energies below $2$ keV (center of mass frame). 
A possible explanation is a correction to the ion energy due to an electron screening potential. 
In comparison to neutron yield predictions with a series of values for the electron screening potential,
they estimate $U_{e}=1000\pm250$ eV in their experiments.

Accordingly, in this work, the calculations are performed for electron screening potentials of
$100, 400, 750, 1000$ eV and $1250$ eV where the corresponding electron Debye length $\lambda_{De}$ in Eq.~\eqref{eq:Ue} are equal to $1.4\times10^{-9}, 3.6\times10^{-10}, 1.9\times10^{-10}, 1.4\times10^{-10} $ and $ 1.1\times10^{-10}$ cm, respectively.

\subsection{Non-resonant data and extrapolation to astrophysical energies}\label{subsec:non-resonant}
For the extrapolation of data points to zero energy, laboratory cross-sections must first be corrected for electron screening~\cite{assenbaum1987effects} before being used in rate calculations. This correction becomes significant when $E/U_{e}\leq100$. In stellar environments, nuclei are surrounded by a dense electron gas, which reduces Coulomb repulsion and facilitates the penetration of the Coulomb barrier. As a result, cross-sections are enhanced compared to those between bare nuclei. This stellar screening effect can be described using, for example, the Debye-Hückel theory of dilute electrolyte solutions~\cite{Debye1923}. Consequently, the connection between stellar and laboratory conditions necessitates the use of bare-nucleus cross-sections.

The enhancement factor $f\left(E\right)$  as a result of the electron screening is determined as~\cite{pinesPRC2020}  
\begin{equation}  
	f\left(E\right)=  \frac{\sigma_{scr.}\left(E\right)}{\sigma\left(E\right)}=\frac{\sigma\left(E+U_{e}\right)}{\sigma\left(E\right)},  \label{eq:enhancement-fac}  
\end{equation}  
where $\sigma_{scr.}\left(E\right)$ and $\sigma\left(E\right)$ represent the cross-sections for screened and bare nuclei, respectively. The former is measured experimentally, while the latter can be derived theoretically. For $U_{e}\ll E$, Eq.~\eqref{eq:enhancement-fac} can be approximated as~\cite{ANGULO19993,DESCOUVEMONT2004203}  
\begin{eqnarray}  
	f\left(E\right) & \simeq\exp\left(\pi\eta U_{e}/E\right) & \,\,\,\,\,\,U_{e}\ll E.\label{eq:enhancement-fac-1}  
\end{eqnarray}  

The screened nuclear reaction cross-section is expressed as~\cite{CzerskiPRC2022, Czerski_2004}  
\begin{eqnarray}  
	\sigma_{scr.}\left(E\right) & = & \frac{1}{\sqrt{E\left(E+U_{e}\right)}}S\left(E\right)\exp\left(-2\pi\eta\sqrt{\frac{E}{E+U_{e}}}\right).\label{eq:sigma-screen}  
\end{eqnarray}

Early measurements of screening energies for heavy metals yielded values around $300$ eV~\cite{Czerski_EPL2001}, exceeding theoretical predictions by approximately a factor of three and significantly surpassing the value of $U_{e}=20\pm5$ eV observed for gaseous targets~\cite{raiola2002enhanced,Jirohta2004, greife1995oppenheimer}. Therefore, results are also presented for screening energies of $100$ and $400$ eV, corresponding to the lower (theoretical) and upper (experimental) limits, as done by Czerski in Ref.~\cite{CzerskiPRC2022}.

In Ref.~\cite{ANGULO19993}, non-resonant data points (from one or multiple works) are represented as $S$-factor data points, and the $S$-factor data are fitted using a polynomial of the form  
\begin{equation}  
	S\left(E\right)\simeq\sum_{i=0}^{N}\mathcal{S}_{i}E^{i}.\label{eq:fit-func-NACRE}  
\end{equation}  
The degree of the polynomial, $N$, is selected to optimize the fit quality, with typical values being $N=2$ or $3$. The coefficients $\mathcal{S}_{i}$ for this polynomial are directly adopted from Ref.~\cite{ANGULO19993} for the $^{2}\textrm{H}\left(d,p\right)^{3}\textrm{H}$ and $^{2}\textrm{H}\left(d,n\right)^{3}\textrm{He}$ reactions and are listed in Table~\ref{tab:Linear-fit-coefficients} as part of the NACRE compilation. These coefficients were derived from a $\chi^{2}$-fit, where individual data points were weighted by their errors, as detailed in Ref.~\cite{ANGULO19993}.

\begin{table}
		\centering
	\begin{tabular}{cccc}
		\hline 
		Compilation &  & $p\:^{3}\textrm{H}$ & $n\:^{3}\textrm{He}$\tabularnewline
		\hline 
		\multirow{3}{*}{NACRE} & $\mathcal{S}_{0}$ & $56.0$ & $55.0$\tabularnewline
		& $\mathcal{S}_{1}$ & $0.204$ & $0.308$\tabularnewline
		& $\mathcal{S}_{2}$ & $-0.0251\times10^{-3}$ & $-0.094\times10^{-3}$\tabularnewline
			\hline 
		\multirow{3}{*}{LEFCS} & $\mathcal{S}_{0}$ & $55.49\left(46\right)$ & $53.76\left(61\right)$\tabularnewline
		& $\mathcal{S}_{1}$ & $0.09482\left(540\right)$ & $0.1623\left(72\right)$\tabularnewline
		& $\mathcal{S}_{2}$ & $0$ & $0$\tabularnewline
		\hline 
	\end{tabular}
	\caption{The $S$ function's fit coefficients, $\mathcal{S}_{0} $ (in keV.b), 
		$\mathcal{S}_{1}$ (in b) and
		$ \mathcal{S}_{2} $ (in  keV$^{-1}$ b)  for the data set, NACRE~\cite{ANGULO19993} and from the measurement at the Los Alamos low-energy fusion cross-section facility (LEFCS)~\cite{BrownPRC1391}. 
		The corresponding fitted function is given by the Eq.~\eqref{eq:fit-func-NACRE}. 
		\label{tab:Linear-fit-coefficients}}
\end{table}

The temperature range over which the reaction rate remains valid depends on the highest energy $E_{d}$ at which the linear fit to $S$ is performed. If the fit's validity is assumed to end at the highest energy ($E_{d}=120$ keV), the integrand of the reactivity integral in Eq.~\eqref{eq:reaction-rate} is applicable only for $kT=0-6$ keV, or equivalently, $0-1.4\: T_{9}$~\cite{BrownPRC1391}.  

\section{Numerical results} \label{sec:numerical-result}
\begin{figure*}[htp]
	\centering
	\includegraphics[scale=0.8]{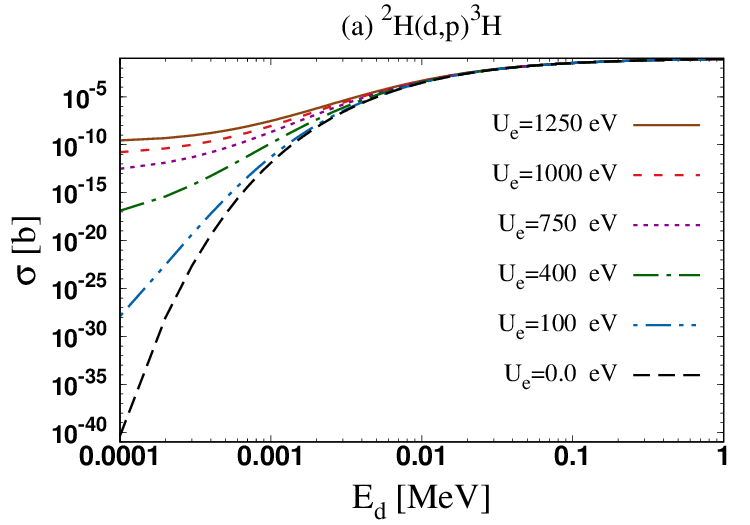} \includegraphics[scale=0.8]{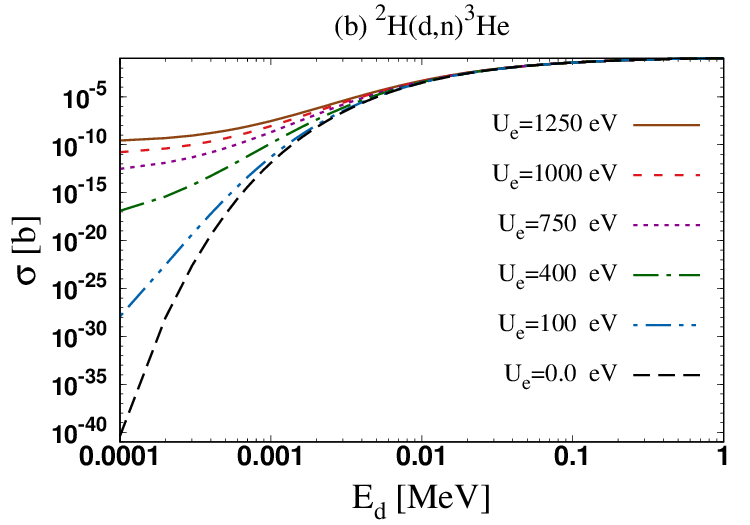} 
	\caption{
		The cross-sections of (a) $ ^{2}\textrm{H}\left(d,p\right) ^{3}\textrm{H}$ and (b) $^{2} \textrm{H} \left(d,n\right) ^{3}\textrm{He}$ reactions along deuteron bombarding energy, $E_{d}$ for 
		$U_{e} = 1250$ eV (solid brown line), $U_{e} = 1000$ eV (short dashed red line), $U_{e} = 750$ eV (dotted magenta line) from Ref.~\cite{Schenkel2019} and for 
		$U_{e} = 400$ eV (dash-dotted green line), $U_{e} = 100$ eV (dash-dot-dotted blue line) from Ref.~\cite{CzerskiPRC2022}
		and without considering the screening effects, i.e., $U_{e} = 0.0$ (dashed black line).
		\label{fig:sigma}}	
\end{figure*}

The cross sections of the $^{2}\textrm{H}\left(d,p\right)^{3}\textrm{H}$ and $^{2}\textrm{H}\left(d,n\right)^{3}\textrm{He}$ reactions, as derived from Eq.~\eqref{eq:sigma-screen}, respectively, are presented in Fig.~\ref{fig:sigma}: (a) and  (b)  versus the deuteron bombarding energy, $E_{d}$. These curves are evaluated for screening energies of $U_{e} = 1250, 1000, 750$ eV and $U_{e} = 400, 100$ eV. The first three values are predicted in Ref.~\cite{Schenkel2019}, while the fourth and fifth values are suggested in Ref.~\cite{CzerskiPRC2022} as the lower (theoretical) and upper (experimental) limits, respectively. For comparison, results without screening effects, i.e., $U_{e} = 0.0$ (represented by a dashed black line), are also included. The latter case is found to be in good agreement with the curves obtained by Bosch-Hale~\cite{Bosch1992} for the D-D reactions, although the employed S-functions differ.

Screening energies for light-ion fusion reactions typically range from tens to hundreds of eV. As illustrated in Fig.~\ref{fig:sigma}, these energies are negligible at high energies but become increasingly significant below 0.01 MeV (10 keV) in the center-of-mass frame.

\begin{figure*}[hbt!]
	\centering
	\includegraphics[scale=0.8]{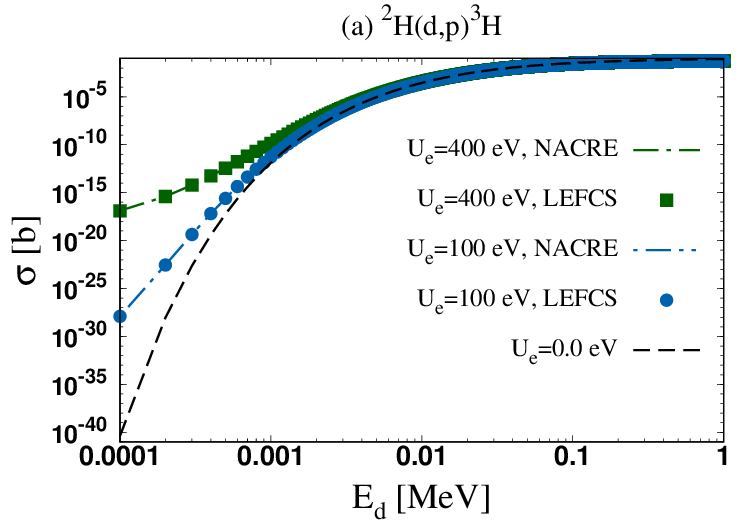} \includegraphics[scale=0.8]{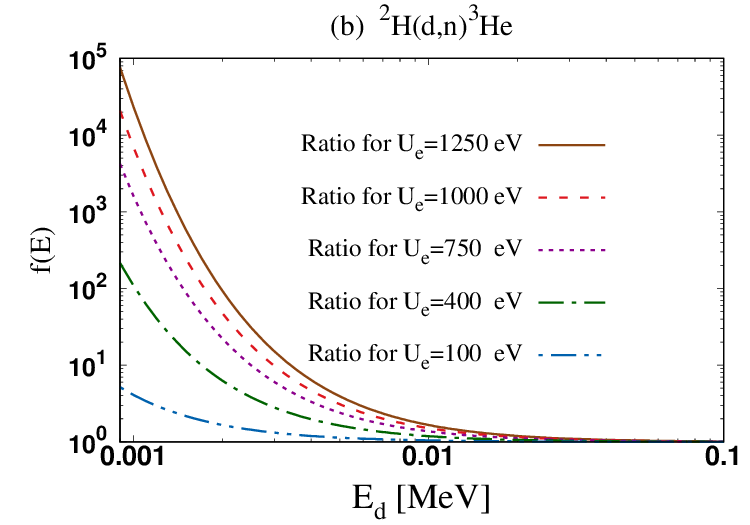} 
	\caption{
		(a) The cross-sections of the $ ^{2}\textrm{H}\left(d,p\right) ^{3}\textrm{H}$ reaction, derived from the S-factor in the NACRE compilation (represented by lines) and LEFCS measurements (represented by data points), are shown for screening energies of $U_{e} = 400$ and $100$ eV.  
		(b) The ratio of the calculated fusion cross-section for screening potentials relative to a bare nucleus, defined as the enhancement factor in Eq.~\ref{eq:enhancement-fac}, is presented. These ratios are computed for $U_{e} = 1250, 1000, 750$ eV and $U_{e} = 400, 100$ eV with the NACRE methodology.
		\label{fig:compareddp}
	}	
\end{figure*}

In Ref.~\cite{CzerskiPRC2022}, the nonresonant and resonant cross sections for the $^{2}\textrm{H}\left(d,\gamma\right)^{4}\textrm{He}$ reaction are presented, with screening energies of 100 and 400 eV, corresponding to the lower (theoretical) and upper (experimental) limit values, respectively. It should be noted that the fitted S-factor function for this reaction in the nonresonant case differs somewhat from those of the other two reactions ($^{2}\textrm{H}\left(d,p\right)^{3}\textrm{H}$ and $^{2}\textrm{H}\left(d,n\right)^{3}\textrm{He}$)~\cite{ANGULO19993}. Even slight differences in these functions can lead to more pronounced variations in the cross-section at very low energies. However, the overall behavior of the cross-section, as shown in Figure 5 of Ref.~\cite{CzerskiPRC2022}, is consistent with the results presented in Figs.~\ref{fig:ddp-ReaRat-fit} and ~\ref{fig:compareddp}: (b) for energies exceeding 0.1 keV or 0.0001 MeV.

	%
As discussed in the last paragraph, some studies, such as Ref.~\cite{CzerskiPRC2022}, utilize the D-D cross sections measured by Brown and Jarmie~\cite{BrownPRC1391}. Therefore, a comparison between the D-D reaction cross sections derived from the fitted S-factor in the NACRE compilation and those measured by Brown and Jarmie~\cite{BrownPRC1391} is warranted. It must be emphasized, however, that the results of Ref.~\cite{BrownPRC1391} were incorporated into the NACRE data fitting process.

The differential cross sections for the $^{2}\textrm{H}\left(d,p\right)^{3}\textrm{H}$ and $^{2}\textrm{H}\left(d,n\right)^{3}\textrm{He}$ reactions were measured by Brown and Jarmie at 11 deuteron bombarding energies ranging from 20 to 117 keV at the Los Alamos low-energy fusion cross-section facility (LEFCS)~\cite{BrownPRC1391}. A linear least-squares fit was applied to the measured S-function data, expressed as $ S = \mathcal{S}_{0} + \mathcal{S}_{1} E_{d} $, where $E_{d}$ represents the lab deuteron bombarding energy in keV. The fitted parameters for $\mathcal{S}_{0}$ (the zero-energy intercepts) and $\mathcal{S}_{1}$ for the two reactions are provided in Table~\ref{tab:Linear-fit-coefficients}, labeled as the LEFCS compilation. Brown and Jarmie asserted that their results significantly improved the accuracy over prior measurements, enabling the most reliable low-temperature reactivities to date. This is corroborated in Fig.~\ref{fig:compareddp}: (a), where the cross-sections derived from the NACRE and LEFCS parameterizations for screening energies of $U_{e} = 400$ and $100$ eV exhibit good agreement within the displayed energy range.

Additionally, Fig.~\ref{fig:compareddp}: (b) displays the ratio of the calculated cross-section of the $^{2} \textrm{H} \left(d,n\right) ^{3}\textrm{He}$ for screening potentials relative to a bare nucleus, defined as the enhancement factor in Eq.~\ref{eq:enhancement-fac}. These enhancement factors are computed for $U_{e} = 1250, 1000, 750$ eV and $U_{e} = 400, 100$ eV.
The present results for \(U_{e}\ll E\) are well described by the approximation given in Eq.~\eqref{eq:enhancement-fac-1}.
 Figure 5 of Ref.~\cite{Berlinguette2019RevisitingTC} illustrates this enhancement factor for $ U_{e} = 400 $ eV, demonstrating its significance in the energy range from 1 to approximately 10 keV. At around 1 keV, Berlinguette et al. approximated the enhancement factor to be about $10^{4}$, whereas our analysis, as shown in Fig.~\ref{fig:compareddp}: (b) for $ U_{e} = 400 $ eV, estimates it to be approximately 100, that there is a gross inconsistency.

The discrepancy could be rooted in the definitions of bound and plasma electron screening, as discussed in Ref.~\cite{Casey2017}. Screening by bound electrons (sometimes referred to as laboratory screening) is particularly important at low energies (i.e., stellar energies) for accelerator-based measurements, requiring model-dependent corrections to obtain S-factors relevant to astrophysical phenomena. The model commonly employed is the same as that given by Eq.~\eqref{eq:enhancement-fac-1}. 

Electron screening of stellar interior plasmas is markedly different from that of terrestrial accelerator experiments, since, for stellar fusion plasmas, the cross-section enhancements depend on plasma properties such as density and temperature, as the screening of the Coulomb barrier is effected by the surrounding plasma electrons. A model commonly used to estimate the enhancement due to plasma electron screening was developed by Salpeter \cite{salpeter1954electron} and is
\begin{equation}
f(E)=\frac{\sigma_{s}(E)}{\sigma(E)}=\exp(\Lambda),
\end{equation}
where the plasma parameter, \(\Lambda = U/(kT)\), is defined as the Coulomb potential energy \(U\) divided by the thermal kinetic energy \(kT\).

The reaction rates, calculated with screening (W/ Scr.) and without screening (W/O Scr.) effects as described in Section~\ref{sec:general-formula}, are tabulated in Tables~\ref{tab:ddp_100_400} and~\ref{tab:ddn_100_400} as a function of $T_{9}$. The temperature steps align with those used in Refs.~\cite{ANGULO19993, DESCOUVEMONT2004203}. The rates are expressed in units of $\textrm{cm}^{3} \textrm{mol}^{-1} \textrm{s}^{-1}$. The column labeled \textit{ratio} displays the ratios between rates with and without screening effects, while the column labeled \textit{NACRE} compares the rates from this work (without screening) to those proposed in Ref.~\cite{ANGULO19993}. The $^{2}\textrm{H}\left(d,p\right)^{3}\textrm{H}$ and $^{2}\textrm{H}\left(d,n\right)^{3}\textrm{He}$ reaction rates are computed for $U_{e} = 100$ and $400$ eV, with results provided in Tables~\ref{tab:ddp_100_400} and~\ref{tab:ddn_100_400}, respectively.

Reaction rate results are omitted for values below $N_{A}\left\langle \sigma v\right\rangle \leq 10^{-25}$ in $\textrm{cm}^{3} \textrm{mol}^{-1} \textrm{s}^{-1}$. As noted in Ref.~\cite{ANGULO19993}, the compilation includes only rates sufficiently large to ensure the target lifetime is shorter than the age of the Universe ($15 \times 10^{9}$ years), corresponding to a minimum temperature of $T_{9} = 0.001$.

A comparison of the $^{2}\textrm{H}\left(d,p\right)^{3}\textrm{H}$ reaction rates for $U_{e} = 100$ and $400$ eV (Table~\ref{tab:ddp_100_400}) reveals a modest enhancement at low $T_{9}$ for $U_{e} = 100$ eV (approximately a factor of 3 at $T_{9} = 0.001$), while a more pronounced increase is observed for $U_{e} = 400$ eV (approximately a factor of 84 at $T_{9} = 0.001$). This enhancement diminishes with rising temperature, eventually converging with the unscreened reaction rate. As anticipated, screening effects are most significant at low energies and temperatures. A similar trend is observed for the $^{2}\textrm{H}\left(d,n\right)^{3}\textrm{He}$ reaction rates (Table~\ref{tab:ddn_100_400}).

To benchmark and a quick validity check of the numerical calculations, the results are compared with the rates from Ref.~\cite{ANGULO19993}. The \textit{NACRE} column in Tables~\ref{tab:ddp_100_400} and~\ref{tab:ddn_100_400} demonstrates excellent agreement between the two datasets.

In Table~\ref{tab:neVsTe}, electron densities $n_{e}$ are presented as a function of temperature $T_{9}$ for screening potentials $U_{e} = 100$ and $400$ eV, as derived from Eq.~\eqref{eq:neVsTe}. The selected $T_{9}$ values correspond to regions where the reaction rates (Tables~\ref{tab:ddp_100_400} and~\ref{tab:ddn_100_400}) exhibit significant magnitudes.

In an ideal plasma, the number of particles per Debye sphere, 
\begin{equation}
	N_{D}\equiv n_{e}\frac{4\pi}{3}\lambda_{D}^{3}\gg1,
\end{equation}
is much greater than unity. Classical plasma theory assumes $N_{D} \gg 1$, emphasizing bulk effects over particle collisions. However, for $N_{D} \leq 1$, screening effects diminish, and particle dynamics become collision-dominated.

\textit{Dense metallic deuterium plasmas}- 
The phase diagram in Figure 4 of Ref.~\cite{ZaghooPRL2019} illustrates the parameter space for various Fermi systems, plotting temperature against carrier number density on a log-log scale. Metallic deuterium plasma (reflectance $\sim 0.7$) forms at approximately $T \sim 11$ eV and carrier density $n \sim 10^{24} \textrm{cm}^{-3}$. The Debye length for this plasma is estimated as $\lambda_{D} \sim 2.5 \times 10^{-9}$ cm, using the relation:
\begin{equation}
	\lambda_{D}=743\left(\frac{T_{e}}{eV}\right)^{1/2}\left(\frac{n_{e}}{cm^{-3}}\right)^{-1/2} \textrm{cm}.
\end{equation}
For dense metallic deuterium plasmas, the screening potential energy is approximately $U_{e} \sim 60$ eV (Eq.~\eqref{eq:Ue}). The corresponding reaction rates are tabulated in Table~\ref{tab:ddp_60}, showing a rate enhancement of about 1.9 at $T_{9} = 0.001$ ($10^{6}$ K).

Moreover, as mentioned in Subsection~\ref{subsec:electronScreen}, the reaction rates are calculated with the values of the screening potentials $U_{e}=750,1000$ and $1250$ eV, which are predicted by Schenkel et al.~\cite{Schenkel2019,Berlinguette2019RevisitingTC} and, respectively, the numerical results are presented in Tables~\ref{tab:ddn_750},~\ref{tab:ddn_1000} and ~\ref{tab:ddn_1250} for the $^{2} \textrm{H} \left(d,n\right) ^{3}\textrm{He}$ reaction rates.

Also, the reaction rates for the $ ^{2}\textrm{H}\left(d,p\right) ^{3}\textrm{H}$ by these values of electron screening potential,
i.e., $U_{e}=750,1000$ and $1250$ eV are given in Tables~\ref{tab:ddp_750},~\ref{tab:ddp_1000} and ~\ref{tab:ddp_1250}, respectively.
These data can be useful for future measurements,
because the branch ratio between the  $p$ + T  and  $n$ + $^{3}$He  channels is going to be determined 
with future implementation of a proton detector~\cite{Schenkel2019}.

\begin{figure*}[hbt!]
	\centering
	\includegraphics[scale=0.8]{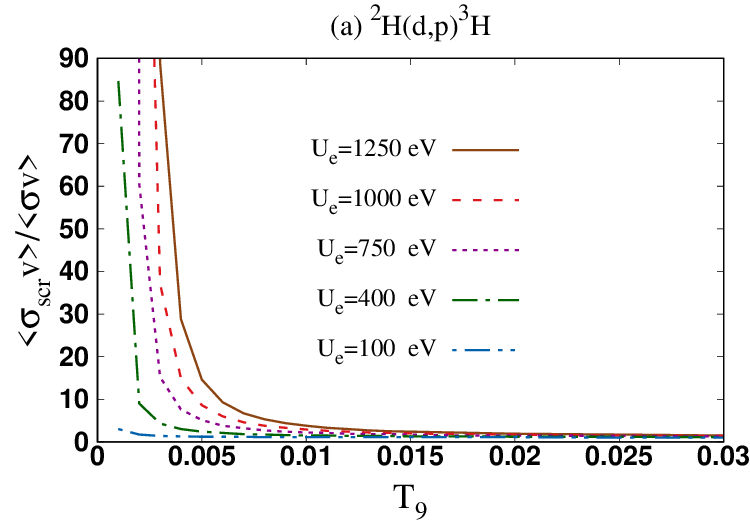} \includegraphics[scale=0.8]{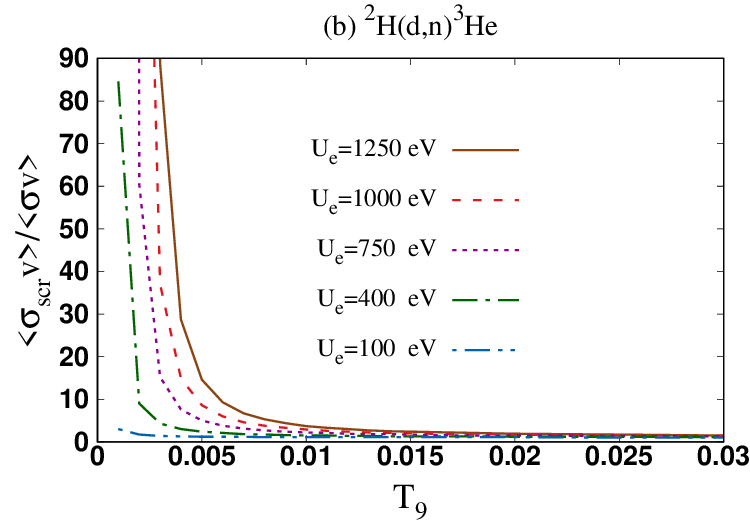} 
	\caption{
		The obtained ratios of the reaction rates with screening  and without screening effects are depicted here as functions of temperature, $T_{9}$, for (a) $ ^{2}\textrm{H}\left(d,p\right) ^{3}\textrm{H}$ and (b) $^{2} \textrm{H} \left(d,n\right) ^{3}\textrm{He}$ reactions. Actually, these are the columns labeled \textit{ratio} in Tables~\ref{tab:ddp_100_400},~\ref{tab:ddp_750},~\ref{tab:ddp_1000},~\ref{tab:ddp_1250} and Tables~\ref{tab:ddn_100_400},~\ref{tab:ddn_750},~\ref{tab:ddn_1000},~\ref{tab:ddn_1250} 
		for $U_{e} = 1250$ eV (solid brown line), $U_{e} = 1000$ eV (short dashed red line), $U_{e} = 750$ eV (dotted magenta line) from Ref.~\cite{Schenkel2019} and for $U_{e} = 400$ eV (dash-dotted green line), $U_{e} = 100$ eV (dash-dot-dotted blue line) from Ref.~\cite{CzerskiPRC2022}.
		\label{fig:reacRatddpddn}
		}	
\end{figure*}

\begin{figure*}[hbt!]
	\centering
	\includegraphics[scale=1.0]{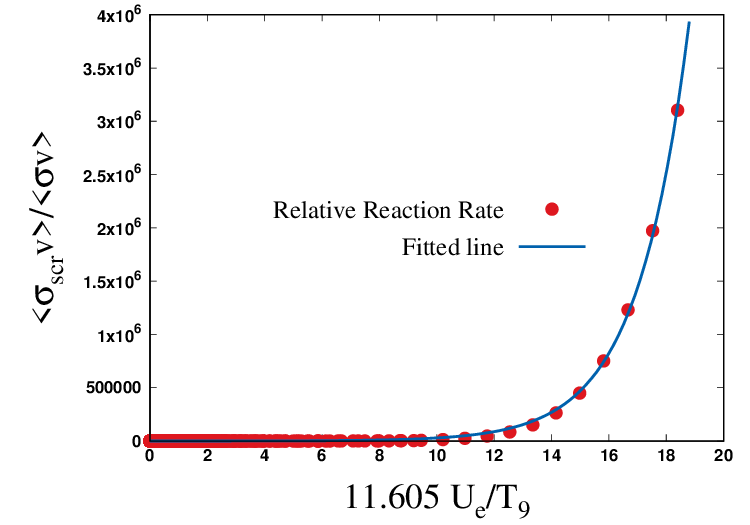}
	\caption{
		The data points (filled red circles) show the relative reaction rate, i.e., the left hand side of the Eq.~\eqref{eq:relative_ReaRat}, as function of $11.605 \: {U_{e}}/{T_{9}}$ for
		$ ^{2}\textrm{H}\left(d,p\right) ^{3}\textrm{H}$ and $^{2} \textrm{H} \left(d,n\right) ^{3}\textrm{He}$ reactions. 
		It should be noted that the units of $U_{e}$ are expressed in MeV.
		This ratio is almost the same for both cases.
		The solid (blue) line shows the obtained fit function as given by the right-hand side of Eq.~\eqref{eq:relative_ReaRat}. 
		\label{fig:ddp-ReaRat-fit}}	
\end{figure*}

To further elucidate the physical concept of changes in reaction rates with variations in screening energy, this behavior is illustrated in Fig.~\ref{fig:reacRatddpddn}. The ratios of the reaction rates, both with screening and without screening effects, are presented in Fig.~\ref{fig:reacRatddpddn}  for the following reactions: (a) $^{2}\textrm{H}\left(d,p\right) ^{3}\textrm{H}$ and (b) $^{2} \textrm{H} \left(d,n\right) ^{3}\textrm{He}$, plotted as functions of temperature, $T_{9}$.  
These ratios correspond to the columns labeled \textit{ratio} in Tables~\ref{tab:ddp_100_400},~\ref{tab:ddp_750},~\ref{tab:ddp_1000},~\ref{tab:ddp_1250} and Tables~\ref{tab:ddn_100_400},~\ref{tab:ddn_750},~\ref{tab:ddn_1000},~\ref{tab:ddn_1250}. The data were derived for screening energies of $U_{e} = 1250, 1000, 750$ eV, as reported in Ref.~\cite{Schenkel2019}, and for $U_{e} = 400, 100$ eV, as reported in Ref.~\cite{CzerskiPRC2022}.  
All of these results confirm that the screening energy, $U_{e}$, serves as a parameter for quantifying the influence of electronic screening on the collision. Notably, an increase in the screening energy is found to enhance the probability of a fusion reaction at lower energies~\cite{Berlinguette2019RevisitingTC, chen2025electrochemical}.

Finally, to demonstrate the dependence of the reaction rate on the screening potential $ U_{e} $ and temperature $ T_{9} $, the ratio of the reaction rate with the screening potential, $ \left\langle \sigma_{scr.} v \right\rangle $, to the reaction rate without the screening potential, $ \left\langle \sigma v \right\rangle $, is analyzed. These reaction rates are obtained by solving the integral in Eq.~\ref{eq:reaction-rate} as a function of $ U_{e} / (k_{B}T) \equiv 11.605 \: U_{e}/T_{9} $, where the relationship $k_{B}T = 0.08617\: T_{9}$ MeV is used; and the results are illustrated in Fig.~\ref{fig:ddp-ReaRat-fit}.     

In this figure, the data points represent the ratio $ \left\langle \sigma_{scr.} v \right\rangle / \left\langle \sigma v \right\rangle $ as a function of the defined parameter $ 11.605 \: U_{e}/T_{9} $. To determine this parameter, for each $ T_{9} $ in the range [0.001, 10.000] (as provided in Table~\ref{tab:ddn_750}), the values of $ U_{e} $ are iterated over by steps of 50 eV within the range [10, 1300] eV. 

Subsequently, the results are fitted with an analytical exponential function, and it is found that the changes in the reaction rate can be described by the following numerical formula:  
\begin{equation}
	\frac{\left\langle \sigma_{scr.} v\right\rangle}{\left\langle \sigma v\right\rangle }\equiv\frac{\int_{0}^{\infty}\sigma_{s}\left(E\right)E\exp\left(-11.605\:E/T_{9}\right)dE}{\int_{0}^{\infty}\sigma\left(E\right)E\exp\left(-11.605\:E/T_{9}\right)dE} = \exp\left(4.70 +6.50{\times10^{-6}}U_{e}/{T_{9}}\right). \label{eq:relative_ReaRat}
\end{equation}
This behavior is nearly identical for both the $ ^{2}\textrm{H}(d,p)^{3}\textrm{H} $ and $ ^{2}\textrm{H}(d,n)^{3}\textrm{He} $ reactions. The similarity between the $ n $ and $ p $ exit channels is expected, as both share the same Coulomb barrier in the entrance channel. 


\section{Summary and conclusions\label{sec:Summary-and-conclusions}} 
The Maxwellian-averaged thermonuclear reaction rates for the $^{2}\textrm{H}\left(d,p\right) ^{3}\textrm{H}$ and $^{2} \textrm{H} \left(d,n\right) ^{3}\textrm{He}$ reactions were calculated, with screening effects taken into consideration. In highly screened environments, such as when hot fusion fuel interacts with lattice nuclei, electron screening may significantly influence tunneling through the Coulomb barrier. Under terrestrial conditions, the simulation of D-D reactions in a metallic deuterium plasma could potentially enhance fusion rates at lower temperatures. 

In this work, the effects of electron screening are phenomenologically described, with screening energy ($U_{e}$) values of 100, 400, 750, 1000, and 1250 eV being utilized, rather than a novel theoretical approach being introduced. 

The laboratory cross-sections were first corrected for electron screening using S-factor functions, which were derived by fitting low-energy data for the $^{2}\textrm{H}\left(d,p\right) ^{3}\textrm{H}$ and $^{2} \textrm{H} \left(d,n\right) ^{3}\textrm{He}$ reactions. For this purpose, a fitted S-factor model based on the NACRE compilation was employed. Subsequently, the integration of Eq.~\eqref{eq:reaction-rate} was performed numerically to compute the Maxwellian-averaged thermonuclear reaction rates.

The corresponding rates, relevant to astrophysical plasmas at temperatures ranging from $10^{6}$ K to $1.3 \times 10^{8}$ K, were tabulated for specified screening energies. These screening energies were motivated by the lower theoretical value of $100$ eV and upper experimental limits of $400$, $750$, $1000$, and $1250$ eV. Additionally, the obtained rates were compared with cases involving no screening and the original NACRE results.  

The numerical results indicated a significant enhancement in the reaction rates for both D-D reactions when the screening energy exceeded $400$ eV. This suggests that higher reaction rates at lower temperatures could be achieved using metallic deuterium plasma.  

Finally, the dependence of the reaction rate on the screening potential, $U_{e}$, was investigated by calculating the ratio $\left\langle \sigma_{scr.} v\right\rangle / \left\langle \sigma v\right\rangle$ as a function of $U_{e}/ \left(k_{B}T\right) = 11.605 \: {U_{e}}/{T_{9}}$. The results were fitted using an analytical function, and the numerical analysis revealed that this ratio could be accurately described by the term $\exp\left(4.70 +6.50{\times10^{-6}}U_{e}/{T_{9}}\right)$ for both $^{2}\textrm{H}\left(d,p\right) ^{3}\textrm{H}$ and $^{2} \textrm{H} \left(d,n\right) ^{3}\textrm{He}$ reactions.  

In conclusion, the screening potential was found to play a significant role in the reaction rate, which depends on temperature. The findings of this work may assist in the design of future fusion experiments for the desired temperature range.  


\section*{Acknowledgement}
The author gratefully acknowledges Dr. S. M. Khorashadizadeh for guidance and technical consultation regarding plasma physics.
\bibliography{Refs.bib}

\begin{table}[hbt!]
		\centering
	\scriptsize 
	\setlength\extrarowheight{-4pt}
	\caption{
		The $ ^{2}\textrm{H}\left(d,p\right) ^{3}\textrm{H}$ reaction rates ($ \textrm{cm}^{3} \textrm{mol}^{-1} \textrm{ s}^{-1} $) with screening (W/ Scr.) and without screening (W/O Scr.).
		The corresponding electron Debye length $\lambda_{De}$ (Eq.~\eqref{eq:Ue}) are equal to $1.4\times10^{-9} $ and $ 3.6\times10^{-10}$ cm for screening potentials $U_{e}=100 $ and $400 $ eV, respectively.		
		\label{tab:ddp_100_400}}	
	\begin{tabular}{ccccc|cccc}
		\hline
		& \multicolumn{4}{c}{ $U_{e}= 100$ eV }    &\multicolumn{4}{c}{$U_{e}= 400$ eV} \\
		\cline{2-5} \cline{6-9}
		$T_{9}$ & W/Scr. & W/O Scr. & ratio &  NACRE 
		& W/ Scr. & W/O Scr. & ratio &  NACRE \\
		\hline
		0.001 &  4.32E-08 &  1.42E-08 & 3.0 &1.0 & 1.20E-06 &  1.42E-08 &84.7 &1.0\\
		0.002 &  1.02E-04 &  5.88E-05 & 1.7 &1.0 & 5.32E-04 &  5.88E-05 & 9.0 &1.0\\
		0.003 &  4.65E-03 &  3.23E-03 & 1.4 &1.0 & 1.39E-02 &  3.23E-03 & 4.3 &1.0\\
		0.004 &  5.22E-02 &  3.99E-02 & 1.3 &1.0 & 1.18E-01 &  3.99E-02 & 3.0 &1.0\\
		0.005 &  2.92E-01 &  2.36E-01 & 1.2 &1.0 & 5.60E-01 &  2.36E-01 & 2.4 &1.0\\
		0.006 &  1.09E+00 &  9.09E-01 & 1.2 &1.0 & 1.87E+00 &  9.09E-01 & 2.1 &1.0\\
		0.007 &  3.10E+00 &  2.66E+00 & 1.2 &1.0 & 4.91E+00 &  2.66E+00 & 1.8 &1.0\\
		0.008 &  7.34E+00 &  6.43E+00 & 1.1 &1.0 & 1.10E+01 &  6.43E+00 & 1.7 &1.0\\
		0.009 &  1.52E+01 &  1.35E+01 & 1.1 &1.0 & 2.17E+01 &  1.35E+01 & 1.6 &1.0\\
		0.010 &  2.85E+01 &  2.56E+01 & 1.1 &1.0 & 3.92E+01 &  2.56E+01 & 1.5 &1.0\\
		0.011 &  4.92E+01 &  4.47E+01 & 1.1 &1.0 & 6.57E+01 &  4.47E+01 & 1.5 &1.0\\
		0.012 &  7.97E+01 &  7.31E+01 & 1.1 &1.0 & 1.04E+02 &  7.31E+01 & 1.4 &1.0\\
		0.013 &  1.23E+02 &  1.13E+02 & 1.1 &1.0 & 1.57E+02 &  1.13E+02 & 1.4 &1.0\\
		0.014 &  1.81E+02 &  1.68E+02 & 1.1 &1.0 & 2.27E+02 &  1.68E+02 & 1.4 &1.0\\
		0.015 &  2.57E+02 &  2.40E+02 & 1.1 &1.0 & 3.18E+02 &  2.40E+02 & 1.3 &1.0\\
		0.016 &  3.55E+02 &  3.33E+02 & 1.1 &1.0 & 4.33E+02 &  3.33E+02 & 1.3 &1.0\\
		0.018 &  6.27E+02 &  5.93E+02 & 1.1 &1.0 & 7.47E+02 &  5.93E+02 & 1.3 &1.0\\
		0.020 &  1.02E+03 &  9.71E+02 & 1.1 &1.0 & 1.19E+03 &  9.71E+02 & 1.2 &1.0\\
		0.025 &  2.70E+03 &  2.60E+03 & 1.0 &1.0 & 3.06E+03 &  2.60E+03 & 1.2 &1.0\\
		0.030 &  5.65E+03 &  5.47E+03 & 1.0 &1.0 & 6.26E+03 &  5.47E+03 & 1.1 &1.0\\
		0.040 &  1.64E+04 &  1.60E+04 & 1.0 &1.0 & 1.77E+04 &  1.60E+04 & 1.1 &1.0\\
		0.050 &  3.48E+04 &  3.42E+04 & 1.0 &1.0 & 3.69E+04 &  3.42E+04 & 1.1 &1.0\\
		0.060 &  6.14E+04 &  6.05E+04 & 1.0 &1.0 & 6.45E+04 &  6.05E+04 & 1.1 &1.0\\
		0.070 &  9.64E+04 &  9.51E+04 & 1.0 &1.0 & 1.00E+05 &  9.51E+04 & 1.1 &1.0\\
		0.080 &  1.39E+05 &  1.38E+05 & 1.0 &1.0 & 1.45E+05 &  1.38E+05 & 1.0 &1.0\\
		0.090 &  1.90E+05 &  1.88E+05 & 1.0 &1.0 & 1.96E+05 &  1.88E+05 & 1.0 &1.0\\
		0.100 &  2.48E+05 &  2.46E+05 & 1.0 &1.0 & 2.55E+05 &  2.46E+05 & 1.0 &1.0\\
		0.110 &  3.13E+05 &  3.10E+05 & 1.0 &1.0 & 3.21E+05 &  3.10E+05 & 1.0 &1.0\\
		0.120 &  3.83E+05 &  3.81E+05 & 1.0 &1.0 & 3.93E+05 &  3.81E+05 & 1.0 &1.0\\
		0.130 &  4.60E+05 &  4.57E+05 & 1.0 &1.0 & 4.70E+05 &  4.57E+05 & 1.0 &1.0\\
		\hline
	\end{tabular}
\end{table}
\begin{table}[hbt!]
		\centering
	\scriptsize 
	\setlength\extrarowheight{-4pt}
	\caption{
		The $^{2} \textrm{H} \left(d,n\right) ^{3}\textrm{He}$ reaction rates ($ \textrm{cm}^{3} \textrm{mol}^{-1} \textrm{ s}^{-1} $). 
		\label{tab:ddn_100_400}}
	\begin{tabular}{ccccc|cccc}	
		\hline
		& \multicolumn{4}{c}{ $U_{e}= 100$ eV }    &\multicolumn{4}{c}{$U_{e}= 400$ eV} \\
		\cline{2-5} \cline{6-9}
		$T_{9}$ & W/ Scr. & W/O Scr. & ratio &  NACRE 
		& W/ Scr. & W/O Scr. & ratio &  NACRE \\
		\hline
		0.001 &  4.26E-08 &  1.40E-08 &3.0 &1.0 &  1.18E-06 &  1.40E-08 &84.6 &1.0\\
		0.002 &  1.01E-04 &  5.80E-05 &1.7 &1.0 &  5.24E-04 &  5.80E-05 & 9.0 &1.0\\
		0.003 &  4.59E-03 &  3.19E-03 &1.4 &1.0 &  1.37E-02 &  3.19E-03 & 4.3 &1.0 \\
		0.004 &  5.16E-02 &  3.94E-02 &1.3 &1.0 &  1.17E-01 &  3.94E-02 & 3.0 &1.0\\
		0.005 &  2.89E-01 &  2.33E-01 &1.2 &1.0 &  5.54E-01 &  2.33E-01 & 2.4 &1.0\\
		0.006 &  1.08E+00 &  9.01E-01 &1.2 &1.0 &  1.85E+00 &  9.01E-01 & 2.1 &1.0\\
		0.007 &  3.07E+00 &  2.64E+00 &1.2 &1.0 &  4.87E+00 &  2.64E+00 & 1.8 &1.0\\
		0.008 &  7.28E+00 &  6.38E+00 &1.1 &1.0 &  1.09E+01 &  6.38E+00 & 1.7 &1.0 \\
		0.009 &  1.51E+01 &  1.34E+01 &1.1 &1.0 &  2.16E+01 &  1.34E+01 & 1.6 &1.0 \\
		0.010 &  2.83E+01 &  2.55E+01 &1.1 &1.0 &  3.89E+01 &  2.55E+01 & 1.5 &1.0\\
		0.011 &  4.89E+01 &  4.45E+01 &1.1 &1.0 &  6.53E+01 &  4.45E+01 & 1.5 &1.0\\
		0.012 &  7.94E+01 &  7.28E+01 &1.1 &1.0 &  1.03E+02 &  7.28E+01 & 1.4 &1.0\\
		0.013 &  1.22E+02 &  1.13E+02 &1.1 &1.0 &  1.56E+02 &  1.13E+02 & 1.4 &1.0\\
		0.014 &  1.80E+02 &  1.68E+02 &1.1 &1.0 &  2.26E+02 &  1.68E+02 & 1.3 &1.0\\
		0.015 &  2.57E+02 &  2.40E+02 &1.1 &1.0 &  3.17E+02 &  2.40E+02 & 1.3 &1.0\\
		0.016 &  3.55E+02 &  3.33E+02 &1.1 &1.0 &  4.32E+02 &  3.33E+02 & 1.3 &1.0 \\
		0.018 &  6.27E+02 &  5.93E+02 &1.1 &1.0 &  7.46E+02 &  5.93E+02 & 1.3 &1.0 \\
		0.020 &  1.02E+03 &  9.72E+02 &1.1 &1.0 &  1.20E+03 &  9.72E+02 & 1.2 &1.0\\
		0.025 &  2.72E+03 &  2.61E+03 &1.0 &1.0 &  3.07E+03 &  2.61E+03 & 1.2 &1.0 \\
		0.030 &  5.69E+03 &  5.51E+03 &1.0 &1.0 &  6.31E+03 &  5.51E+03 & 1.1 &1.0\\
		0.040 &  1.66E+04 &  1.62E+04 &1.0 &1.0 &  1.79E+04 &  1.62E+04 & 1.1 &1.0\\
		0.050 &  3.54E+04 &  3.48E+04 &1.0 &1.0 &  3.76E+04 &  3.48E+04 & 1.1 &1.0 \\
		0.060 &  6.28E+04 &  6.18E+04 &1.0 &1.0 &  6.59E+04 &  6.18E+04 & 1.1 &1.0 \\
		0.070 &  9.89E+04 &  9.77E+04 &1.0 &1.0 &  1.03E+05 &  9.77E+04 & 1.1 &1.0 \\
		0.080 &  1.44E+05 &  1.42E+05 &1.0 &1.0 &  1.49E+05 &  1.42E+05 & 1.0 &1.0\\
		0.090 &  1.97E+05 &  1.95E+05 &1.0 &1.0 &  2.03E+05 &  1.95E+05 & 1.0 &1.0\\
		0.100 &  2.58E+05 &  2.55E+05 &1.0 &1.0 &  2.65E+05 &  2.55E+05 & 1.0 &1.0 \\
		0.110 &  3.26E+05 &  3.23E+05 &1.0 &1.0 &  3.34E+05 &  3.23E+05 & 1.0 &1.0 \\
		0.120 &  4.01E+05 &  3.98E+05 &1.0 &1.0 &  4.10E+05 &  3.98E+05 & 1.0 &1.0\\
		0.130 &  4.82E+05 &  4.79E+05 &1.0 &1.0 &  4.92E+05 &  4.79E+05 & 1.0 &1.0\\
		\hline
	\end{tabular}
\end{table}
\begin{table}[hbt!]
		\centering
	\scriptsize
	\caption{
		The electron density $n_{e}$ (in cm$^{-3}$) versus temperature $T_{9}$ for two fixed value of screening potential $ U_{e}=100 $ and $400$ eV as given by Eq.~\eqref{eq:neVsTe}. \label{tab:neVsTe}}.
	\begin{tabular} 
		{p{0.08\textwidth}>{\centering}p{0.08\textwidth}>{\centering}p{0.08\textwidth}>{\centering}p{0.08\textwidth}>{\centering}p{0.08\textwidth}>{\centering}p{0.08\textwidth} 			p{0.08\textwidth}>{\centering}p{0.08\textwidth}>{\centering}p{0.08\textwidth}>{\centering}p{0.08\textwidth}>{\centering\arraybackslash}p{0.08\textwidth} }
		\hline 
		  $T_{9}$ & 0.001 & 0.002 & 0.003 & 0.004 & 0.005 & 0.006 & 0.007 & 0.008 & 0.009 & 0.010\\		
		\hline 		
		 $U_{e}=$100 &  2.30E+25 &  4.59E+25 &  6.89E+25 &  9.19E+25 &  1.15E+26 &  1.38E+26 &  1.61E+26 &  1.84E+26 &  2.07E+26 &  2.30E+26\\
		 $U_{e}=$400 &  3.67E+26 &  7.35E+26 &  1.10E+27 &  1.47E+27 &  1.84E+27 &  2.20E+27 &  2.57E+27 &  2.94E+27 &  3.31E+27 &  3.67E+27\\	
		\hline
	\end{tabular}
\end{table}
\begin{table}[hbt!]
		\centering
	\scriptsize 
	\setlength\extrarowheight{-4pt}
	\caption{
		The $ ^{2}\textrm{H}\left(d,p\right) ^{3}\textrm{H}$ reaction rates ($ \textrm{cm}^{3} \textrm{mol}^{-1} \textrm{ s}^{-1} $) for  $U_{e}= 60$ eV; 
		and $\lambda_{De}=2.4\times10^{-9} $ cm. \label{tab:ddp_60}}	
	\begin{tabular}{cccc|cccc}
		\hline
		$T_{9}$ & W/ Scr. & W/O Scr. & ratio &  
		$T_{9}$ & W/ Scr. & W/O Scr. & ratio \\
		\hline
		$10^{-5}$&  3.72E-49 &  1.35E-74 &  2.76E+25 &  0.009 &  1.45E+01 &  1.35E+01 &1.1 \\
		$10^{-4}$&  2.68E-26 &  2.91E-29 &  9.20E+02 &  0.010 &  2.73E+01 &  2.56E+01 &1.1 \\
		0.001 &  2.77E-08 &  1.42E-08 & 1.9  & 0.011 &  4.73E+01 &  4.47E+01 &1.1 \\
		0.002 &  8.17E-05 &  5.88E-05 & 1.4  & 0.012 &  7.69E+01 &  7.31E+01 &1.1 \\
		0.003 &  4.02E-03 &  3.23E-03 & 1.2  & 0.013 &  1.19E+02 &  1.13E+02 &1.0 \\
		0.004 &  4.68E-02 &  3.99E-02 & 1.2  & 0.014 &  1.76E+02 &  1.68E+02 &1.0 \\
		0.005 &  2.68E-01 &  2.36E-01 & 1.1  & 0.015 &  2.50E+02 &  2.40E+02 &1.0 \\
		0.006 &  1.01E+00 &  9.09E-01 & 1.1  & 0.016 &  3.46E+02 &  3.33E+02 &1.0 \\
		0.007 &  2.91E+00 &  2.66E+00 & 1.1  & 0.018 &  6.13E+02 &  5.93E+02 &1.0 \\
		0.008 &  6.96E+00 &  6.43E+00 & 1.1  & 0.020 &  1.00E+03 &  9.71E+02 &1.0 \\
		\hline
	\end{tabular}
\end{table}
\begin{table}[hbt!]
		\centering
	\scriptsize 
	\setlength\extrarowheight{-4pt}
	\caption{
		The $^{2} \textrm{H} \left(d,n\right) ^{3}\textrm{He}$ reaction rates ($ \textrm{cm}^{3} \textrm{mol}^{-1} \textrm{ s}^{-1} $) for  $U_{e}= 750$ eV;
		and $\lambda_{De}=1.9\times10^{-10} $ cm. 
		\label{tab:ddn_750}}	
	\begin{tabular}{ccccc|ccccc}
		\hline
		$T_{9}$ & W/ Scr. & W/O Scr. & ratio &  $n_{e}$ &
		$T_{9}$ & W/ Scr. & W/O Scr. & ratio &  $n_{e}$ \\
		\hline
		0.001 &  5.09E-05 &  1.40E-08 & 3639.8 &  1.29E+15 & 0.140 &  5.94E+05 &  5.66E+05 &1.0 &  1.81E+17\\
		0.002 &  3.50E-03 &  5.80E-05 &   60.4 &  2.58E+15 & 0.150 &  6.88E+05 &  6.59E+05 &1.0 &  1.94E+17\\
		0.003 &  4.85E-02 &  3.19E-03 &   15.2 &  3.88E+15 & 0.160 &  7.88E+05 &  7.56E+05 &1.0 &  2.07E+17\\
		0.004 &  3.00E-01 &  3.94E-02 &    7.6 &  5.17E+15 & 0.180 &  1.00E+06 &  9.65E+05 &1.0 &  2.33E+17\\
		0.005 &  1.18E+00 &  2.33E-01 &    5.0 &  6.46E+15 & 0.200 &  1.23E+06 &  1.19E+06 &1.0 &  2.58E+17\\
		0.006 &  3.45E+00 &  9.01E-01 &    3.8 &  7.75E+15 & 0.250 &  1.85E+06 &  1.81E+06 &1.0 &  3.23E+17\\
		0.007 &  8.31E+00 &  2.64E+00 &    3.1 &  9.04E+15 & 0.300 &  2.53E+06 &  2.48E+06 &1.0 &  3.88E+17\\
		0.008 &  1.74E+01 &  6.38E+00 &    2.7 &  1.03E+16 & 0.350 &  3.25E+06 &  3.19E+06 &1.0 &  4.52E+17\\
		0.009 &  3.26E+01 &  1.34E+01 &    2.4 &  1.16E+16 & 0.400 &  3.99E+06 &  3.93E+06 &1.0 &  5.17E+17\\
		0.010 &  5.64E+01 &  2.55E+01 &    2.2 &  1.29E+16 & 0.450 &  4.75E+06 &  4.69E+06 &1.0 &  5.81E+17\\
		0.011 &  9.15E+01 &  4.45E+01 &    2.1 &  1.42E+16 & 0.500 &  5.52E+06 &  5.46E+06 &1.0 &  6.46E+17\\
		0.012 &  1.41E+02 &  7.28E+01 &    1.9 &  1.55E+16 & 0.600 &  7.07E+06 &  7.01E+06 &1.0 &  7.75E+17\\
		0.013 &  2.07E+02 &  1.13E+02 &    1.8 &  1.68E+16 & 0.700 &  8.63E+06 &  8.57E+06 &1.0 &  9.04E+17\\
		0.014 &  2.94E+02 &  1.68E+02 &    1.8 &  1.81E+16 & 0.800 &  1.02E+07 &  1.01E+07 &1.0 &  1.03E+18\\
		0.015 &  4.05E+02 &  2.40E+02 &    1.7 &  1.94E+16 & 0.900 &  1.17E+07 &  1.16E+07 &1.0 &  1.16E+18\\
		0.016 &  5.43E+02 &  3.33E+02 &    1.6 &  2.07E+16 & 1.000 &  1.32E+07 &  1.31E+07 &1.0 &  1.29E+18\\
		0.018 &  9.13E+02 &  5.93E+02 &    1.5 &  2.33E+16 & 1.250 &  1.68E+07 &  1.68E+07 &1.0 &  1.61E+18\\
		0.020 &  1.43E+03 &  9.72E+02 &    1.5 &  2.58E+16 & 1.500 &  2.02E+07 &  2.02E+07 &1.0 &  1.94E+18\\
		0.025 &  3.55E+03 &  2.61E+03 &    1.4 &  3.23E+16 & 1.750 &  2.35E+07 &  2.34E+07 &1.0 &  2.26E+18\\
		0.030 &  7.10E+03 &  5.51E+03 &    1.3 &  3.88E+16 & 2.000 &  2.65E+07 &  2.65E+07 &1.0 &  2.58E+18\\
		0.040 &  1.96E+04 &  1.62E+04 &    1.2 &  5.17E+16 & 2.500 &  3.21E+07 &  3.20E+07 &1.0 &  3.23E+18\\
		0.050 &  4.03E+04 &  3.48E+04 &    1.2 &  6.46E+16 & 3.000 &  3.70E+07 &  3.70E+07 &1.0 &  3.88E+18\\
		0.060 &  6.98E+04 &  6.18E+04 &    1.1 &  7.75E+16 & 3.500 &  4.13E+07 &  4.13E+07 &1.0 &  4.52E+18\\
		0.070 &  1.08E+05 &  9.77E+04 &    1.1 &  9.04E+16 & 4.000 &  4.51E+07 &  4.51E+07 &1.0 &  5.17E+18\\
		0.080 &  1.55E+05 &  1.42E+05 &    1.1 &  1.03E+17 & 5.000 &  5.14E+07 &  5.14E+07 &1.0 &  6.46E+18\\
		0.090 &  2.11E+05 &  1.95E+05 &    1.1 &  1.16E+17 & 6.000 &  5.60E+07 &  5.60E+07 &1.0 &  7.75E+18\\
		0.100 &  2.74E+05 &  2.55E+05 &    1.1 &  1.29E+17 & 7.000 &  5.92E+07 &  5.92E+07 &1.0 &  9.04E+18\\
		0.110 &  3.44E+05 &  3.23E+05 &    1.1 &  1.42E+17 & 8.000 &  6.11E+07 &  6.11E+07 &1.0 &  1.03E+19\\
		0.120 &  4.21E+05 &  3.98E+05 &    1.1 &  1.55E+17 & 9.000 &  6.20E+07 &  6.20E+07 &1.0 &  1.16E+19\\
		0.130 &  5.05E+05 &  4.79E+05 &    1.1 &  1.68E+17 &10.000 &  6.18E+07 &  6.18E+07 &1.0 &  1.29E+19\\
		\hline
	\end{tabular}
\end{table}
\begin{table}[hbt!]
		\centering
	\scriptsize 
	\setlength\extrarowheight{-4pt}
	\caption{
		The $^{2} \textrm{H} \left(d,n\right) ^{3}\textrm{He}$ reaction rates ($ \textrm{cm}^{3} \textrm{mol}^{-1} \textrm{ s}^{-1} $) for  $U_{e}= 1000$ eV; 
		and $\lambda_{De}=1.4\times10^{-10} $ cm. \label{tab:ddn_1000}}	
	\begin{tabular}{ccccc|ccccc}
		\hline
		$T_{9}$ & W/ Scr. & W/O Scr. & ratio &  $n_{e}$ &
		$T_{9}$ & W/ Scr. & W/O Scr. & ratio &  $n_{e}$ \\
		\hline
		0.001 &  5.81E-04 &  1.40E-08 &41530.7 &  2.30E+15 & 0.140 &  6.03E+05 &  5.66E+05 &1.1 &  3.22E+17\\
		0.002 &  1.32E-02 &  5.80E-05 &  227.6 &  4.59E+15 & 0.150 &  6.99E+05 &  6.59E+05 &1.1 &  3.44E+17\\
		0.003 &  1.18E-01 &  3.19E-03 &   37.0 &  6.89E+15 & 0.160 &  7.99E+05 &  7.56E+05 &1.1 &  3.67E+17\\
		0.004 &  5.86E-01 &  3.94E-02 &   14.9 &  9.19E+15 & 0.180 &  1.01E+06 &  9.65E+05 &1.0 &  4.13E+17\\
		0.005 &  2.01E+00 &  2.33E-01 &    8.6 &  1.15E+16 & 0.200 &  1.24E+06 &  1.19E+06 &1.0 &  4.59E+17\\
		0.006 &  5.38E+00 &  9.01E-01 &    6.0 &  1.38E+16 & 0.250 &  1.87E+06 &  1.81E+06 &1.0 &  5.74E+17\\
		0.007 &  1.21E+01 &  2.64E+00 &    4.6 &  1.61E+16 & 0.300 &  2.55E+06 &  2.48E+06 &1.0 &  6.89E+17\\
		0.008 &  2.42E+01 &  6.38E+00 &    3.8 &  1.84E+16 & 0.350 &  3.26E+06 &  3.19E+06 &1.0 &  8.04E+17\\
		0.009 &  4.37E+01 &  1.34E+01 &    3.3 &  2.07E+16 & 0.400 &  4.01E+06 &  3.93E+06 &1.0 &  9.19E+17\\
		0.010 &  7.34E+01 &  2.55E+01 &    2.9 &  2.30E+16 & 0.450 &  4.77E+06 &  4.69E+06 &1.0 &  1.03E+18\\
		0.011 &  1.16E+02 &  4.45E+01 &    2.6 &  2.53E+16 & 0.500 &  5.54E+06 &  5.46E+06 &1.0 &  1.15E+18\\
		0.012 &  1.75E+02 &  7.28E+01 &    2.4 &  2.76E+16 & 0.600 &  7.09E+06 &  7.01E+06 &1.0 &  1.38E+18\\
		0.013 &  2.53E+02 &  1.13E+02 &    2.2 &  2.99E+16 & 0.700 &  8.65E+06 &  8.57E+06 &1.0 &  1.61E+18\\
		0.014 &  3.54E+02 &  1.68E+02 &    2.1 &  3.22E+16 & 0.800 &  1.02E+07 &  1.01E+07 &1.0 &  1.84E+18\\
		0.015 &  4.81E+02 &  2.40E+02 &    2.0 &  3.44E+16 & 0.900 &  1.17E+07 &  1.16E+07 &1.0 &  2.07E+18\\
		0.016 &  6.38E+02 &  3.33E+02 &    1.9 &  3.67E+16 & 1.000 &  1.32E+07 &  1.31E+07 &1.0 &  2.30E+18\\
		0.018 &  1.05E+03 &  5.93E+02 &    1.8 &  4.13E+16 & 1.250 &  1.68E+07 &  1.68E+07 &1.0 &  2.87E+18\\
		0.020 &  1.63E+03 &  9.72E+02 &    1.7 &  4.59E+16 & 1.500 &  2.02E+07 &  2.02E+07 &1.0 &  3.44E+18\\
		0.025 &  3.93E+03 &  2.61E+03 &    1.5 &  5.74E+16 & 1.750 &  2.35E+07 &  2.34E+07 &1.0 &  4.02E+18\\
		0.030 &  7.73E+03 &  5.51E+03 &    1.4 &  6.89E+16 & 2.000 &  2.65E+07 &  2.65E+07 &1.0 &  4.59E+18\\
		0.040 &  2.08E+04 &  1.62E+04 &    1.3 &  9.19E+16 & 2.500 &  3.21E+07 &  3.20E+07 &1.0 &  5.74E+18\\
		0.050 &  4.23E+04 &  3.48E+04 &    1.2 &  1.15E+17 & 3.000 &  3.70E+07 &  3.70E+07 &1.0 &  6.89E+18\\
		0.060 &  7.27E+04 &  6.18E+04 &    1.2 &  1.38E+17 & 3.500 &  4.13E+07 &  4.13E+07 &1.0 &  8.04E+18\\
		0.070 &  1.12E+05 &  9.77E+04 &    1.1 &  1.61E+17 & 4.000 &  4.52E+07 &  4.51E+07 &1.0 &  9.19E+18\\
		0.080 &  1.60E+05 &  1.42E+05 &    1.1 &  1.84E+17 & 5.000 &  5.14E+07 &  5.14E+07 &1.0 &  1.15E+19\\
		0.090 &  2.16E+05 &  1.95E+05 &    1.1 &  2.07E+17 & 6.000 &  5.60E+07 &  5.60E+07 &1.0 &  1.38E+19\\
		0.100 &  2.80E+05 &  2.55E+05 &    1.1 &  2.30E+17 & 7.000 &  5.92E+07 &  5.92E+07 &1.0 &  1.61E+19\\
		0.110 &  3.51E+05 &  3.23E+05 &    1.1 &  2.53E+17 & 8.000 &  6.12E+07 &  6.11E+07 &1.0 &  1.84E+19\\
		0.120 &  4.29E+05 &  3.98E+05 &    1.1 &  2.76E+17 & 9.000 &  6.20E+07 &  6.20E+07 &1.0 &  2.07E+19\\
		0.130 &  5.13E+05 &  4.79E+05 &    1.1 &  2.99E+17 &10.000 &  6.18E+07 &  6.18E+07 &1.0 &  2.30E+19\\
		\hline
	\end{tabular}
\end{table}
\begin{table}[hbt!]
		\centering
	\scriptsize 
	\setlength\extrarowheight{-4pt}
	\caption{
		The $^{2} \textrm{H} \left(d,n\right) ^{3}\textrm{He}$ reaction rates ($ \textrm{cm}^{3} \textrm{mol}^{-1} \textrm{ s}^{-1} $) for  $U_{e}= 1250$ eV; 
		and $\lambda_{De}=1.1\times10^{-10} $ cm. \label{tab:ddn_1250}}	
	\begin{tabular}{ccccc|ccccc}
		\hline
		$T_{9}$ & W/ Scr. & W/O Scr. & ratio &  $n_{e}$ &
		$T_{9}$ & W/ Scr. & W/O Scr. & ratio &  $n_{e}$ \\
		\hline
		0.001 &  4.63E-03 &  1.40E-08 & 3.3E+05 &  3.59E+15 & 0.140 &  6.13E+05 &  5.66E+05 &1.1 &  5.02E+17\\
		0.002 &  4.72E-02 &  5.80E-05 &   813.7 &  7.18E+15 & 0.150 &  7.09E+05 &  6.59E+05 &1.1 &  5.38E+17\\
		0.003 &  2.83E-01 &  3.19E-03 &    88.7 &  1.08E+16 & 0.160 &  8.10E+05 &  7.56E+05 &1.1 &  5.74E+17\\
		0.004 &  1.13E+00 &  3.94E-02 &    28.7 &  1.44E+16 & 0.180 &  1.02E+06 &  9.65E+05 &1.1 &  6.46E+17\\
		0.005 &  3.40E+00 &  2.33E-01 &    14.6 &  1.79E+16 & 0.200 &  1.25E+06 &  1.19E+06 &1.1 &  7.18E+17\\
		0.006 &  8.34E+00 &  9.01E-01 &     9.3 &  2.15E+16 & 0.250 &  1.88E+06 &  1.81E+06 &1.0 &  8.97E+17\\
		0.007 &  1.77E+01 &  2.64E+00 &     6.7 &  2.51E+16 & 0.300 &  2.56E+06 &  2.48E+06 &1.0 &  1.08E+18\\
		0.008 &  3.36E+01 &  6.38E+00 &     5.3 &  2.87E+16 & 0.350 &  3.28E+06 &  3.19E+06 &1.0 &  1.26E+18\\
		0.009 &  5.85E+01 &  1.34E+01 &     4.4 &  3.23E+16 & 0.400 &  4.03E+06 &  3.93E+06 &1.0 &  1.44E+18\\
		0.010 &  9.54E+01 &  2.55E+01 &     3.7 &  3.59E+16 & 0.450 &  4.79E+06 &  4.69E+06 &1.0 &  1.61E+18\\
		0.011 &  1.47E+02 &  4.45E+01 &     3.3 &  3.95E+16 & 0.500 &  5.56E+06 &  5.46E+06 &1.0 &  1.79E+18\\
		0.012 &  2.17E+02 &  7.28E+01 &     3.0 &  4.31E+16 & 0.600 &  7.11E+06 &  7.01E+06 &1.0 &  2.15E+18\\
		0.013 &  3.09E+02 &  1.13E+02 &     2.7 &  4.67E+16 & 0.700 &  8.67E+06 &  8.57E+06 &1.0 &  2.51E+18\\
		0.014 &  4.26E+02 &  1.68E+02 &     2.5 &  5.02E+16 & 0.800 &  1.02E+07 &  1.01E+07 &1.0 &  2.87E+18\\
		0.015 &  5.72E+02 &  2.40E+02 &     2.4 &  5.38E+16 & 0.900 &  1.17E+07 &  1.16E+07 &1.0 &  3.23E+18\\
		0.016 &  7.50E+02 &  3.33E+02 &     2.3 &  5.74E+16 & 1.000 &  1.32E+07 &  1.31E+07 &1.0 &  3.59E+18\\
		0.018 &  1.22E+03 &  5.93E+02 &     2.1 &  6.46E+16 & 1.250 &  1.68E+07 &  1.68E+07 &1.0 &  4.49E+18\\
		0.020 &  1.85E+03 &  9.72E+02 &     1.9 &  7.18E+16 & 1.500 &  2.03E+07 &  2.02E+07 &1.0 &  5.38E+18\\
		0.025 &  4.35E+03 &  2.61E+03 &     1.7 &  8.97E+16 & 1.750 &  2.35E+07 &  2.34E+07 &1.0 &  6.28E+18\\
		0.030 &  8.41E+03 &  5.51E+03 &     1.5 &  1.08E+17 & 2.000 &  2.65E+07 &  2.65E+07 &1.0 &  7.18E+18\\
		0.040 &  2.22E+04 &  1.62E+04 &     1.4 &  1.44E+17 & 2.500 &  3.21E+07 &  3.20E+07 &1.0 &  8.97E+18\\
		0.050 &  4.44E+04 &  3.48E+04 &     1.3 &  1.79E+17 & 3.000 &  3.70E+07 &  3.70E+07 &1.0 &  1.08E+19\\
		0.060 &  7.57E+04 &  6.18E+04 &     1.2 &  2.15E+17 & 3.500 &  4.13E+07 &  4.13E+07 &1.0 &  1.26E+19\\
		0.070 &  1.16E+05 &  9.77E+04 &     1.2 &  2.51E+17 & 4.000 &  4.52E+07 &  4.51E+07 &1.0 &  1.44E+19\\
		0.080 &  1.65E+05 &  1.42E+05 &     1.2 &  2.87E+17 & 5.000 &  5.14E+07 &  5.14E+07 &1.0 &  1.79E+19\\
		0.090 &  2.22E+05 &  1.95E+05 &     1.1 &  3.23E+17 & 6.000 &  5.60E+07 &  5.60E+07 &1.0 &  2.15E+19\\
		0.100 &  2.87E+05 &  2.55E+05 &     1.1 &  3.59E+17 & 7.000 &  5.92E+07 &  5.92E+07 &1.0 &  2.51E+19\\
		0.110 &  3.59E+05 &  3.23E+05 &     1.1 &  3.95E+17 & 8.000 &  6.12E+07 &  6.11E+07 &1.0 &  2.87E+19\\
		0.120 &  4.38E+05 &  3.98E+05 &     1.1 &  4.31E+17 & 9.000 &  6.20E+07 &  6.20E+07 &1.0 &  3.23E+19\\
		0.130 &  5.23E+05 &  4.79E+05 &     1.1 &  4.67E+17 &10.000 &  6.18E+07 &  6.18E+07 &1.0 &  3.59E+19\\
		\hline
	\end{tabular}
\end{table}
\begin{table}[hbt!]
		\centering
	\scriptsize 
	\setlength\extrarowheight{-4pt}
	\caption{
		The $ ^{2}\textrm{H}\left(d,p\right) ^{3}\textrm{H}$ reaction rates ($ \textrm{cm}^{3} \textrm{mol}^{-1} \textrm{ s}^{-1} $) for  $U_{e}= 750$ eV. 
		\label{tab:ddp_750}}	
	\begin{tabular}{ccccc|ccccc}
		\hline
		$T_{9}$ & W/ Scr. & W/O Scr. & ratio &  $n_{e}$ &
		$T_{9}$ & W/ Scr. & W/O Scr. & ratio &  $n_{e}$ \\
		\hline
		0.001 &  5.18E-05 &  1.42E-08 & 3644.7 &  1.29E+15 & 0.140 &  5.65E+05 &  5.38E+05 &1.0 &  1.81E+17\\
		0.002 &  3.56E-03 &  5.88E-05 &   60.5 &  2.58E+15 & 0.150 &  6.53E+05 &  6.25E+05 &1.0 &  1.94E+17\\
		0.003 &  4.92E-02 &  3.23E-03 &   15.2 &  3.88E+15 & 0.160 &  7.46E+05 &  7.15E+05 &1.0 &  2.07E+17\\
		0.004 &  3.04E-01 &  3.99E-02 &    7.6 &  5.17E+15 & 0.180 &  9.41E+05 &  9.08E+05 &1.0 &  2.33E+17\\
		0.005 &  1.19E+00 &  2.36E-01 &    5.1 &  6.46E+15 & 0.200 &  1.15E+06 &  1.11E+06 &1.0 &  2.58E+17\\
		0.006 &  3.49E+00 &  9.09E-01 &    3.8 &  7.75E+15 & 0.250 &  1.71E+06 &  1.67E+06 &1.0 &  3.23E+17\\
		0.007 &  8.39E+00 &  2.66E+00 &    3.2 &  9.04E+15 & 0.300 &  2.32E+06 &  2.27E+06 &1.0 &  3.88E+17\\
		0.008 &  1.75E+01 &  6.43E+00 &    2.7 &  1.03E+16 & 0.350 &  2.95E+06 &  2.90E+06 &1.0 &  4.52E+17\\
		0.009 &  3.29E+01 &  1.35E+01 &    2.4 &  1.16E+16 & 0.400 &  3.60E+06 &  3.55E+06 &1.0 &  5.17E+17\\
		0.010 &  5.68E+01 &  2.56E+01 &    2.2 &  1.29E+16 & 0.450 &  4.26E+06 &  4.21E+06 &1.0 &  5.81E+17\\
		0.011 &  9.21E+01 &  4.47E+01 &    2.1 &  1.42E+16 & 0.500 &  4.92E+06 &  4.87E+06 &1.0 &  6.46E+17\\
		0.012 &  1.41E+02 &  7.31E+01 &    1.9 &  1.55E+16 & 0.600 &  6.24E+06 &  6.18E+06 &1.0 &  7.75E+17\\
		0.013 &  2.08E+02 &  1.13E+02 &    1.8 &  1.68E+16 & 0.700 &  7.54E+06 &  7.49E+06 &1.0 &  9.04E+17\\
		0.014 &  2.95E+02 &  1.68E+02 &    1.8 &  1.81E+16 & 0.800 &  8.82E+06 &  8.77E+06 &1.0 &  1.03E+18\\
		0.015 &  4.06E+02 &  2.40E+02 &    1.7 &  1.94E+16 & 0.900 &  1.01E+07 &  1.00E+07 &1.0 &  1.16E+18\\
		0.016 &  5.44E+02 &  3.33E+02 &    1.6 &  2.07E+16 & 1.000 &  1.13E+07 &  1.13E+07 &1.0 &  1.29E+18\\
		0.018 &  9.15E+02 &  5.93E+02 &    1.5 &  2.33E+16 & 1.250 &  1.43E+07 &  1.42E+07 &1.0 &  1.61E+18\\
		0.020 &  1.43E+03 &  9.71E+02 &    1.5 &  2.58E+16 & 1.500 &  1.71E+07 &  1.70E+07 &1.0 &  1.94E+18\\
		0.025 &  3.54E+03 &  2.60E+03 &    1.4 &  3.23E+16 & 1.750 &  1.97E+07 &  1.97E+07 &1.0 &  2.26E+18\\
		0.030 &  7.06E+03 &  5.47E+03 &    1.3 &  3.88E+16 & 2.000 &  2.22E+07 &  2.22E+07 &1.0 &  2.58E+18\\
		0.040 &  1.93E+04 &  1.60E+04 &    1.2 &  5.17E+16 & 2.500 &  2.69E+07 &  2.68E+07 &1.0 &  3.23E+18\\
		0.050 &  3.96E+04 &  3.42E+04 &    1.2 &  6.46E+16 & 3.000 &  3.11E+07 &  3.11E+07 &1.0 &  3.88E+18\\
		0.060 &  6.83E+04 &  6.05E+04 &    1.1 &  7.75E+16 & 3.500 &  3.51E+07 &  3.50E+07 &1.0 &  4.52E+18\\
		0.070 &  1.06E+05 &  9.51E+04 &    1.1 &  9.04E+16 & 4.000 &  3.87E+07 &  3.87E+07 &1.0 &  5.17E+18\\
		0.080 &  1.51E+05 &  1.38E+05 &    1.1 &  1.03E+17 & 5.000 &  4.52E+07 &  4.52E+07 &1.0 &  6.46E+18\\
		0.090 &  2.04E+05 &  1.88E+05 &    1.1 &  1.16E+17 & 6.000 &  5.10E+07 &  5.10E+07 &1.0 &  7.75E+18\\
		0.100 &  2.64E+05 &  2.46E+05 &    1.1 &  1.29E+17 & 7.000 &  5.60E+07 &  5.60E+07 &1.0 &  9.04E+18\\
		0.110 &  3.31E+05 &  3.10E+05 &    1.1 &  1.42E+17 & 8.000 &  6.04E+07 &  6.04E+07 &1.0 &  1.03E+19\\
		0.120 &  4.03E+05 &  3.81E+05 &    1.1 &  1.55E+17 & 9.000 &  6.44E+07 &  6.44E+07 &1.0 &  1.16E+19\\
		0.130 &  4.82E+05 &  4.57E+05 &    1.1 &  1.68E+17 &10.000 &  6.79E+07 &  6.79E+07 &1.0 &  1.29E+19\\		
		\hline
	\end{tabular}
\end{table}
\begin{table}[hbt!]
		\centering
	\scriptsize 
	\setlength\extrarowheight{-4pt}
	\caption{
		The $ ^{2}\textrm{H}\left(d,p\right) ^{3}\textrm{H}$ reaction rates ($ \textrm{cm}^{3} \textrm{mol}^{-1} \textrm{ s}^{-1} $) for  $U_{e}= 1000$ eV. 
		\label{tab:ddp_1000}}	
	\begin{tabular}{ccccc|ccccc}
		\hline
		$T_{9}$ & W/ Scr. & W/O Scr. & ratio &  $n_{e}$ &
		$T_{9}$ & W/ Scr. & W/O Scr. & ratio &  $n_{e}$ \\
		\hline
		0.001 &  5.91E-04 &  1.42E-08 &41598.9 &  2.30E+15 & 0.140 &  5.75E+05 &  5.38E+05 &1.1 &  3.22E+17\\
		0.002 &  1.34E-02 &  5.88E-05 &  228.0 &  4.59E+15 & 0.150 &  6.63E+05 &  6.25E+05 &1.1 &  3.44E+17\\
		0.003 &  1.20E-01 &  3.23E-03 &   37.1 &  6.89E+15 & 0.160 &  7.56E+05 &  7.15E+05 &1.1 &  3.67E+17\\
		0.004 &  5.94E-01 &  3.99E-02 &   14.9 &  9.19E+15 & 0.180 &  9.53E+05 &  9.08E+05 &1.1 &  4.13E+17\\
		0.005 &  2.03E+00 &  2.36E-01 &    8.6 &  1.15E+16 & 0.200 &  1.16E+06 &  1.11E+06 &1.0 &  4.59E+17\\
		0.006 &  5.44E+00 &  9.09E-01 &    6.0 &  1.38E+16 & 0.250 &  1.73E+06 &  1.67E+06 &1.0 &  5.74E+17\\
		0.007 &  1.23E+01 &  2.66E+00 &    4.6 &  1.61E+16 & 0.300 &  2.34E+06 &  2.27E+06 &1.0 &  6.89E+17\\
		0.008 &  2.44E+01 &  6.43E+00 &    3.8 &  1.84E+16 & 0.350 &  2.97E+06 &  2.90E+06 &1.0 &  8.04E+17\\
		0.009 &  4.41E+01 &  1.35E+01 &    3.3 &  2.07E+16 & 0.400 &  3.62E+06 &  3.55E+06 &1.0 &  9.19E+17\\
		0.010 &  7.40E+01 &  2.56E+01 &    2.9 &  2.30E+16 & 0.450 &  4.28E+06 &  4.21E+06 &1.0 &  1.03E+18\\
		0.011 &  1.17E+02 &  4.47E+01 &    2.6 &  2.53E+16 & 0.500 &  4.94E+06 &  4.87E+06 &1.0 &  1.15E+18\\
		0.012 &  1.76E+02 &  7.31E+01 &    2.4 &  2.76E+16 & 0.600 &  6.26E+06 &  6.18E+06 &1.0 &  1.38E+18\\
		0.013 &  2.55E+02 &  1.13E+02 &    2.2 &  2.99E+16 & 0.700 &  7.56E+06 &  7.49E+06 &1.0 &  1.61E+18\\
		0.014 &  3.56E+02 &  1.68E+02 &    2.1 &  3.22E+16 & 0.800 &  8.84E+06 &  8.77E+06 &1.0 &  1.84E+18\\
		0.015 &  4.83E+02 &  2.40E+02 &    2.0 &  3.44E+16 & 0.900 &  1.01E+07 &  1.00E+07 &1.0 &  2.07E+18\\
		0.016 &  6.40E+02 &  3.33E+02 &    1.9 &  3.67E+16 & 1.000 &  1.13E+07 &  1.13E+07 &1.0 &  2.30E+18\\
		0.018 &  1.06E+03 &  5.93E+02 &    1.8 &  4.13E+16 & 1.250 &  1.43E+07 &  1.42E+07 &1.0 &  2.87E+18\\
		0.020 &  1.63E+03 &  9.71E+02 &    1.7 &  4.59E+16 & 1.500 &  1.71E+07 &  1.70E+07 &1.0 &  3.44E+18\\
		0.025 &  3.92E+03 &  2.60E+03 &    1.5 &  5.74E+16 & 1.750 &  1.97E+07 &  1.97E+07 &1.0 &  4.02E+18\\
		0.030 &  7.68E+03 &  5.47E+03 &    1.4 &  6.89E+16 & 2.000 &  2.22E+07 &  2.22E+07 &1.0 &  4.59E+18\\
		0.040 &  2.06E+04 &  1.60E+04 &    1.3 &  9.19E+16 & 2.500 &  2.69E+07 &  2.68E+07 &1.0 &  5.74E+18\\
		0.050 &  4.16E+04 &  3.42E+04 &    1.2 &  1.15E+17 & 3.000 &  3.11E+07 &  3.11E+07 &1.0 &  6.89E+18\\
		0.060 &  7.12E+04 &  6.05E+04 &    1.2 &  1.38E+17 & 3.500 &  3.51E+07 &  3.50E+07 &1.0 &  8.04E+18\\
		0.070 &  1.09E+05 &  9.51E+04 &    1.1 &  1.61E+17 & 4.000 &  3.87E+07 &  3.87E+07 &1.0 &  9.19E+18\\
		0.080 &  1.55E+05 &  1.38E+05 &    1.1 &  1.84E+17 & 5.000 &  4.53E+07 &  4.52E+07 &1.0 &  1.15E+19\\
		0.090 &  2.09E+05 &  1.88E+05 &    1.1 &  2.07E+17 & 6.000 &  5.10E+07 &  5.10E+07 &1.0 &  1.38E+19\\
		0.100 &  2.70E+05 &  2.46E+05 &    1.1 &  2.30E+17 & 7.000 &  5.60E+07 &  5.60E+07 &1.0 &  1.61E+19\\
		0.110 &  3.38E+05 &  3.10E+05 &    1.1 &  2.53E+17 & 8.000 &  6.05E+07 &  6.04E+07 &1.0 &  1.84E+19\\
		0.120 &  4.11E+05 &  3.81E+05 &    1.1 &  2.76E+17 & 9.000 &  6.44E+07 &  6.44E+07 &1.0 &  2.07E+19\\
		0.130 &  4.90E+05 &  4.57E+05 &    1.1 &  2.99E+17 &10.000 &  6.79E+07 &  6.79E+07 &1.0 &  2.30E+19\\		
		\hline
	\end{tabular}
\end{table}
\begin{table}[hbt!]
		\centering
	\scriptsize 
	\setlength\extrarowheight{-4pt}
	\caption{
		The $ ^{2}\textrm{H}\left(d,p\right) ^{3}\textrm{H}$ reaction rates ($ \textrm{cm}^{3} \textrm{mol}^{-1} \textrm{ s}^{-1} $) for  $U_{e}= 1250$ eV. 
		\label{tab:ddp_1250}}	
	\begin{tabular}{ccccc|ccccc}
		\hline
		$T_{9}$ & W/ Scr. & W/O Scr. & ratio &  $n_{e}$ &
		$T_{9}$ & W/ Scr. & W/O Scr. & ratio &  $n_{e}$ \\
		\hline
		0.001 &  4.71E-03 &  1.42E-08 &3.3E+05 &  3.59E+15 & 0.140 &  5.84E+05 &  5.38E+05 &1.1 &  5.02E+17\\
		0.002 &  4.80E-02 &  5.88E-05 &   815.5 &  7.18E+15 & 0.150 &  6.73E+05 &  6.25E+05 &1.1 &  5.38E+17\\
		0.003 &  2.87E-01 &  3.23E-03 &    88.9 &  1.08E+16 & 0.160 &  7.67E+05 &  7.15E+05 &1.1 &  5.74E+17\\
		0.004 &  1.15E+00 &  3.99E-02 &    28.8 &  1.44E+16 & 0.180 &  9.65E+05 &  9.08E+05 &1.1 &  6.46E+17\\
		0.005 &  3.44E+00 &  2.36E-01 &    14.6 &  1.79E+16 & 0.200 &  1.18E+06 &  1.11E+06 &1.1 &  7.18E+17\\
		0.006 &  8.44E+00 &  9.09E-01 &     9.3 &  2.15E+16 & 0.250 &  1.74E+06 &  1.67E+06 &1.0 &  8.97E+17\\
		0.007 &  1.79E+01 &  2.66E+00 &     6.7 &  2.51E+16 & 0.300 &  2.35E+06 &  2.27E+06 &1.0 &  1.08E+18\\
		0.008 &  3.39E+01 &  6.43E+00 &     5.3 &  2.87E+16 & 0.350 &  2.99E+06 &  2.90E+06 &1.0 &  1.26E+18\\
		0.009 &  5.90E+01 &  1.35E+01 &     4.4 &  3.23E+16 & 0.400 &  3.64E+06 &  3.55E+06 &1.0 &  1.44E+18\\
		0.010 &  9.62E+01 &  2.56E+01 &     3.8 &  3.59E+16 & 0.450 &  4.30E+06 &  4.21E+06 &1.0 &  1.61E+18\\
		0.011 &  1.48E+02 &  4.47E+01 &     3.3 &  3.95E+16 & 0.500 &  4.96E+06 &  4.87E+06 &1.0 &  1.79E+18\\
		0.012 &  2.19E+02 &  7.31E+01 &     3.0 &  4.31E+16 & 0.600 &  6.28E+06 &  6.18E+06 &1.0 &  2.15E+18\\
		0.013 &  3.11E+02 &  1.13E+02 &     2.7 &  4.67E+16 & 0.700 &  7.58E+06 &  7.49E+06 &1.0 &  2.51E+18\\
		0.014 &  4.29E+02 &  1.68E+02 &     2.5 &  5.02E+16 & 0.800 &  8.86E+06 &  8.77E+06 &1.0 &  2.87E+18\\
		0.015 &  5.75E+02 &  2.40E+02 &     2.4 &  5.38E+16 & 0.900 &  1.01E+07 &  1.00E+07 &1.0 &  3.23E+18\\
		0.016 &  7.53E+02 &  3.33E+02 &     2.3 &  5.74E+16 & 1.000 &  1.13E+07 &  1.13E+07 &1.0 &  3.59E+18\\
		0.018 &  1.22E+03 &  5.93E+02 &     2.1 &  6.46E+16 & 1.250 &  1.43E+07 &  1.42E+07 &1.0 &  4.49E+18\\
		0.020 &  1.85E+03 &  9.71E+02 &     1.9 &  7.18E+16 & 1.500 &  1.71E+07 &  1.70E+07 &1.0 &  5.38E+18\\
		0.025 &  4.34E+03 &  2.60E+03 &     1.7 &  8.97E+16 & 1.750 &  1.97E+07 &  1.97E+07 &1.0 &  6.28E+18\\
		0.030 &  8.36E+03 &  5.47E+03 &     1.5 &  1.08E+17 & 2.000 &  2.22E+07 &  2.22E+07 &1.0 &  7.18E+18\\
		0.040 &  2.19E+04 &  1.60E+04 &     1.4 &  1.44E+17 & 2.500 &  2.69E+07 &  2.68E+07 &1.0 &  8.97E+18\\
		0.050 &  4.37E+04 &  3.42E+04 &     1.3 &  1.79E+17 & 3.000 &  3.12E+07 &  3.11E+07 &1.0 &  1.08E+19\\
		0.060 &  7.42E+04 &  6.05E+04 &     1.2 &  2.15E+17 & 3.500 &  3.51E+07 &  3.50E+07 &1.0 &  1.26E+19\\
		0.070 &  1.13E+05 &  9.51E+04 &     1.2 &  2.51E+17 & 4.000 &  3.87E+07 &  3.87E+07 &1.0 &  1.44E+19\\
		0.080 &  1.60E+05 &  1.38E+05 &     1.2 &  2.87E+17 & 5.000 &  4.53E+07 &  4.52E+07 &1.0 &  1.79E+19\\
		0.090 &  2.15E+05 &  1.88E+05 &     1.1 &  3.23E+17 & 6.000 &  5.10E+07 &  5.10E+07 &1.0 &  2.15E+19\\
		0.100 &  2.77E+05 &  2.46E+05 &     1.1 &  3.59E+17 & 7.000 &  5.60E+07 &  5.60E+07 &1.0 &  2.51E+19\\
		0.110 &  3.45E+05 &  3.10E+05 &     1.1 &  3.95E+17 & 8.000 &  6.05E+07 &  6.04E+07 &1.0 &  2.87E+19\\
		0.120 &  4.19E+05 &  3.81E+05 &     1.1 &  4.31E+17 & 9.000 &  6.44E+07 &  6.44E+07 &1.0 &  3.23E+19\\
		0.130 &  4.99E+05 &  4.57E+05 &     1.1 &  4.67E+17 &10.000 &  6.79E+07 &  6.79E+07 &1.0 &  3.59E+19\\
		\hline
	\end{tabular}
\end{table}
\end{document}